\documentclass[fleqn,usenatbib]{mnras}
\usepackage{newtxtext,newtxmath}
\usepackage[T1]{fontenc}
\DeclareRobustCommand{\VAN}[3]{#2}
\let\VANthebibliography\thebibliography
\def\thebibliography{\DeclareRobustCommand{\VAN}[3]{##3}\VANthebibliography}

\usepackage{graphicx}	
\usepackage{amsmath}	
\usepackage{color}
\usepackage{threeparttable}
\usepackage{array,booktabs}
\usepackage{siunitx}
\usepackage{subcaption}

\usepackage{threeparttable}

\usepackage{CJKutf8}

\defcitealias{Toyouchi2025MNRAS}{T25}
\defcitealias{Hirashita2019MNRAS}{H19}

\newcommand{\msun}{{\rm M_\odot}} 
\newcommand{\zsun}{{\rm Z_\odot}} 
\newcommand{\yr}{{\rm yr}} 
\newcommand{\Myr}{{\rm Myr}} 
\newcommand{\Gyr}{{\rm Gyr}} 
\newcommand{\pc}{{\rm pc}} 
\newcommand{\kpc}{{\rm kpc}} 
\newcommand{\kelvin}{{\rm K}} 

\newcommand{\mhfive}{M_{\rm h5}}

\newcommand{\mstar}{M_\ast}

\newcommand{\rmin}{R_{\rm min}}
\newcommand{\rmax}{R_{\rm max}}

\newcommand{\sgmg}{\Sigma_{\rm g}}
\newcommand{\sgms}{\Sigma_\ast}

\newcommand{\sgmsf}{\dot{\Sigma}_{\rm sf}}
\newcommand{\sgmin}{\dot{\Sigma}_{\rm in}}
\newcommand{\sgmout}{\dot{\Sigma}_{\rm out}}
\newcommand{\sgmret}{\dot{\Sigma}_{\rm ret}}
\newcommand{\sgmzr}{\dot{\Sigma}_{{\rm ret},i}}
\newcommand{\mret}{m_{\rm ret}}

\newcommand{\mmax}{m_{\rm \ast,max}}
\newcommand{\mmin}{m_{\rm \ast,min}}

\newcommand{\yd}{y_{\rm d}}

\newcommand{\rhodw}{\rho_{\rm d,w}}
\newcommand{\rhodc}{\rho_{\rm d,c}}

\newcommand{\rhog}{\rho_{\rm g}}
\newcommand{\Tg}{T_{\rm g}}

\newcommand{\Tw}{T_{\rm w}}
\newcommand{\Tc}{T_{\rm c}}
\newcommand{\fcm}{f_{\rm c}^{\rm M}}
\newcommand{\fcv}{f_{\rm c}^{\rm V}}
\newcommand{\rhoc}{\rho_{\rm c}}
\newcommand{\rhow}{\rho_{\rm w}}
\newcommand{\rhoavg}{\overline{\rho}}
\newcommand{\nc}{n_{\rm c}}
\newcommand{\nw}{n_{\rm w}}

\newcommand{\muc}{\mu_{\rm c}}
\newcommand{\muw}{\mu_{\rm w}}

\newcommand{\nIIc}{\dot{n}_{\rm II,c}}
\newcommand{\nIIw}{\dot{n}_{\rm II,w}}
\newcommand{\rhodav}{\overline{\rho}_{\rm d}}

\newcommand{\keff}{\kappa_{\lambda}^{\rm eff}}
\newcommand{\kabs}{\kappa_{\lambda}^{\rm abs}}
\newcommand{\ksca}{\kappa_{\lambda}^{\rm sca}}
\newcommand{\kas}{\kappa_{\lambda}^{\rm abs/sca}}
\newcommand{\Qext}{Q^{\rm ext}_{i}(a,\, \lambda)}
\newcommand{\Qas}{Q^{\rm abs/sca}_{i}(a,\, \lambda)}
\newcommand{\Qabs}{Q^{\rm abs}_{i}}
\newcommand{\Qsca}{Q^{\rm sca}_{i}}

\newcommand{\tauc}{\tau_{\lambda,\, \rm c}}
\newcommand{\tauw}{\tau_{\lambda,\, \rm w}}

\newcommand{\rc}{r_{\rm c}}

\title[Dust growth in high-redshift galaxies]{Grain-size evolution and rapid dust growth in high-redshift galaxies}
\author[D.~Toyouchi et al.]{
Daisuke~Toyouchi$^{1}$\thanks{E-mail: d.toyouchi@gmail.com},
Andrea~Ferrara$^{2}$,
Yurina~Nakazato$^{3}$,
Kosei~Matsumoto$^{4}$,
\newauthor \vspace{1.5em}
~Raffaella~Schneider$^{5,6,7,8}$,
Koki~Otaki$^{9}$
\\
$^{1}$Theoretical Astrophysics, Department of Earth \& Space Science, Graduate School of Science, Osaka University, \\
1-1 Machikaneyama, Toyonaka, Osaka, 560-0043, Japan\\
$^{2}$Scuola Normale Superiore, Piazza dei Cavalieri 7, 50126 Pisa, Italy\\
$^{3}$Center for Computational Astrophysics, Flatiron Institute, 162 5th Avenue, New York, NY 10010, USA\\
$^{4}$Sterrenkundig Observatorium Department of Physics and Astronomy Universiteit Gent, Krijgslaan 281 S9, 9000 Gent, Belgium\\
$^{5}$Dipartimento di Fisica, “Sapienza” Universit\`a di Roma, Piazzale Aldo Moro 5, I-00185 Roma, Italy\\
$^{6}$INAF/Osservatorio Astronomico di Roma, Via Frascati 33, I-00040 Monte Porzio Catone, Italy\\
$^{7}$INFN, Sezione Roma I, Dipartimento di Fisica, “Sapienza” Universit\`a di Roma, Piazzale Aldo Moro 2, I-00185, Roma, Italy\\
$^{8}$Sapienza School for Advanced Studies, Viale Regina Elena 291, I-00161 Roma, Italy\\
$^{9}$Information Technology Center, The University of Tokyo, 6-2-3 Kashiwanoha, Kashiwa, Chiba 277-0882, Japan\\
}

\date{Accepted XXX. Received YYY; in original form ZZZ}

\pubyear{2026}

\begin{document}
\label{firstpage}
\pagerange{\pageref{firstpage}--\pageref{lastpage}}
\maketitle

\begin{abstract}
We present a galaxy evolution model that incorporates grain-size evolution in a multiphase interstellar medium (ISM) to investigate dust attenuation in galaxies at $z \geq 5$.
Our fiducial setup assumes a low dust yield of $\yd = 10^{-4}~\msun$ and a small characteristic size of stellar dust of $a_0 = 0.01~\mu$m, motivated by efficient dust destruction by reverse shocks in dense ISM environments. Our model demonstrates that, even with such low dust yields, massive galaxies with $M_\ast > 10^9~\msun$ reach high dust-to-stellar mass ratios of $M_{\rm d}/M_\ast \sim 10^{-2}$ by $z \sim 7$ because small grains supplied by SNe efficiently serve as seeds for metal accretion in the ISM.
Such rapid dust growth leads to a substantial population of large grains ($a \gtrsim 0.1~\mu{\rm m}$), resulting in relatively flat attenuation curves.
Because dust growth significantly lags behind star formation, the outer regions beyond the half-star-formation-rate radius remain relatively dust poor, allowing a non-negligible fraction of UV photons to escape without strong attenuation.
We further find that dust growth becomes most efficient when the ISM is dominated by cold dense gas but still contains a modest warm component, as the former promotes metal accretion while the latter supplies additional small grains through shattering, thereby further enhancing subsequent grain growth.
In particular, with a cold dense gas fraction of $\sim 90~\%$,
our model predictions become broadly consistent with the dust-to-stellar mass ratios inferred for dust-rich galaxies at $z \sim 7$, as well as the upper limits for blue galaxies at $z \gtrsim 10$.
Self-consistently, the model successfully reproduces the UV luminosity functions observed at both $z = 7$ and $z = 12$.
Overall, this study demonstrates that a physically motivated treatment of grain growth in a multiphase ISM is essential for linking the dust content of high-redshift galaxies to their radiative properties during cosmic dawn.
\end{abstract}

\begin{keywords}
galaxies: evolution -- galaxies: high-redshift -- galaxies: luminosity function, mass function -- galaxies: star formation
\end{keywords}



\section{Introduction}\label{sec:intro}

Dust plays a fundamental role in regulating the observable properties of galaxies. Dust grains absorb ultraviolet (UV) and optical radiation emitted by stars and reprocess it into infrared (IR) emission, thereby shaping the spectral energy distributions (SEDs) of galaxies across a wide range of wavelengths. Dust also affects the thermal and chemical evolution of the interstellar medium (ISM) by catalyzing molecular hydrogen formation and modifying gas cooling processes. Therefore, understanding the origin and evolution of dust is essential for interpreting observations of galaxy evolution, particularly during the early stages of cosmic history.

Recent observations with the \textit{James Webb Space Telescope} (JWST) have dramatically advanced our understanding of dust in galaxies at cosmic dawn. In particular, JWST surveys have revealed a large population of UV-bright galaxies at $z > 10$, often referred to as ``blue monsters,'' 
characterized by extremely blue UV continua with spectral slopes of $\beta \lesssim -2.2$ \citep[e.g.,][]{Topping2022ApJ, Topping2024MNRAS, Cullen2024MNRAS, Morales2024ApJ}.
Such blue SEDs imply that these galaxies experience very weak dust attenuation despite their intense star formation activity. In addition, non-detections of dust continuum suggest extremely low dust-to-stellar mass ratios, $M_{\rm d}/M_\ast \lesssim 10^{-4}$, for many galaxies at $z > 10$ \citep{Fudamoto2024MNRAS, Carniani2025A&A, Mitsuhashi2026ApJ, Bakx2025MNRAS, Bakx2026MNRAS}. These results indicate that dust production and growth may still be highly inefficient during the earliest phases of galaxy evolution.


At the same time, ALMA observations have revealed the opposite aspect of dust evolution during the Epoch of Reionization ($z \sim 6$-8). In particular, the REBELS survey and related studies have discovered a substantial population of dust-rich galaxies at $z \sim 7$, many of which exhibit dust-to-stellar mass ratios comparable to or even higher than those of local star-forming galaxies \citep{Bouwens2022ApJ, Inami2022MNRAS, Barrufet2023MNRAS}. These observations imply that dust can grow rapidly within galaxies on timescales of only a few hundred Myr. Therefore, galaxies at $z > 10$ and those at $z \sim 7$ appear to represent two contrasting phases of dust evolution: one characterized by extremely weak dust attenuation and another by rapid dust enrichment.

Motivated by these discoveries, theoretical studies of dust evolution in galaxy formation have rapidly progressed in recent years \citep[e.g.,][]{Dayal2010MNRAS, Bekki2015MNRAS, Popping2017MNRAS, McKinnon2017MNRAS, Dayal2022MNRAS, Trebitsch2023MNRAS}. 
In particular, the evolution of grain-size distributions is now recognized as a key ingredient because grain growth and destruction timescales strongly depend on grain size, and grain sizes directly determine extinction and attenuation curves. To capture these effects efficiently, many studies adopted the two-size approximation, in which the entire grain population is represented by small and large grains \citep{Hou2017MNRAS, Aoyama2017MNRAS, Gjergo2018MNRAS, Parente2022MNRAS, Parente2023MNRAS, Osman2025arXiv}. More recently, numerical simulations that explicitly follow full grain-size distributions have achieved more physically accurate descriptions of dust growth and radiative properties \citep{McKinnon2018MNRAS, Aoyama2020MNRAS, Matsumoto2024A&A, Matsumoto2026A&A, Choban2026arXiv}. However, such full grain-size calculations are computationally expensive and are therefore generally limited to idealized setups or zoom-in simulations.

Recently, \citet{Narayanan2026OJAp} performed cosmological galaxy formation simulations incorporating full grain-size evolution up to $z \sim 6$, successfully reproducing both UV-bright galaxies at $z > 10$ and dust-rich galaxies at $z \sim 7$.
Their extremely high spatial resolution and detailed analysis enabled them to demonstrate that rapid dust growth proceeds preferentially in dense gas, highlighting the crucial role of inhomogeneous ISM structure in regulating dust evolution at high redshift.
While these simulations represent a major step toward a unified understanding of dust evolution during cosmic dawn, their high computational cost makes it difficult to systematically explore uncertain dust-related parameters, such as the dust yield and characteristic grain size of supernova (SN)-produced dust.

In this context, semi-analytic galaxy evolution models provide a complementary and physically transparent approach \citep{Vijayan2019MNRAS, Hirashita2019MNRAS, Nishida2022MNRAS, Kano2026arXiv}.
Recently, \citet{Parente2026arXiv} incorporated full grain-size evolution into the semi-analytic galaxy formation model L-G{\scriptsize ALAXIES} 2020.
They showed that initially large-grain-dominated dust populations produced by SNe gradually evolve toward MRN-like grain-size distributions \citep{Jones1996ApJ} through ISM processing. 
Furthermore, their model successfully reproduced the cosmic dust mass densities at $z < 2$.
On the other hand, comparisons between such semi-analytic models and observations at $z > 4$ remain limited.
In particular, semi-analytic studies capable of comprehensively investigating the rapid buildup of dust at $z \gtrsim 7$, together with its impact on grain-size distributions, dust attenuation, and the UV/IR radiative properties of galaxies, are still needed.

In this paper, we incorporate a full grain-size evolution model into the semi-analytic galaxy evolution framework developed in our previous study \citep[][hereafter T25]{Toyouchi2025MNRAS}, which successfully reproduced a variety of observed galaxy properties at $z > 5$, including the size evolution of galaxies, the stellar mass--gas-phase metallicity relation, and the cosmic star formation rate (SFR) density.
\citetalias{Toyouchi2025MNRAS} also included a simplified treatment of dust evolution and showed that the rapid dust buildup observed at $z > 7$ can be generally reproduced by assuming a very short grain growth timescale, $\tau_{\rm acc} = 5~{\rm Myr}~(Z/Z_\odot)^{-1}$ \citep[see also][]{Bakx2026MNRAS}.
However, the grain growth timescale should fundamentally depend on both the grain-size distribution and the physical conditions of the ISM.

For this purpose, we explicitly consider a multiphase ISM consisting of cold dense gas and warm diffuse gas.
These phases play distinct roles in dust evolution: the former provides the primary sites of rapid dust growth via metal accretion, while the latter promotes dust destruction and the production of small grains through shattering.
Using this framework, we comprehensively investigate dust production and grain-size evolution in high-redshift galaxies, and examine how evolving grain-size distributions affect observable properties such as the UV and IR luminosity functions.

The paper is organized as follows.
In \S~\ref{sec:methods}, we describe the galaxy evolution model, the treatment of grain-size evolution in a two-phase ISM, and the radiative transfer calculations used to derive dust attenuation and UV/IR luminosities.
In \S~\ref{sec:results}, we present the evolution of dust mass and grain size distributions, along with the resulting extinction and attenuation properties.
In \S~\ref{sec:rad}, we compare the model predictions with the observed UV and IR luminosity functions and discuss the constraints on dust physics in high-$z$ galaxies.
In \S~\ref{sec:discuss}, we discuss the physical implications of our results and the connection to attenuation-free scenarios.
Finally, we summarize our conclusions in \S~\ref{sec:summary}.

Throughout this paper, we use the Planck cosmological parameter sets: $\Omega_{\rm m} = 0.3111$, $\Omega_{\rm \Lambda} = 0.6899$, $\Omega_{\rm b} = 0.0489$, and $h = 0.6766$, and $\sigma_8 = 0.8102$ \citep[][]{Planck2020A&A}.


\section{Methods}\label{sec:methods}

\subsection{Galaxy evolution model}\label{sec:gal_model}

In this section, we briefly summarize the setup of our semi-analytic galaxy evolution model (see \citetalias{Toyouchi2025MNRAS} for more details).
Our model calculates galaxy evolution from $z = 20$ to $z = 5$, corresponding to the cosmic age of $0.18~{\rm Gyr}$ and $1.2~{\rm Gyr}$, respectively.
We assume that galaxies already possess entirely rotation-supported disks \citep[$v/\sigma \gtrsim 1$, e.g.,][]{Parlanti2023A&A, Nelson2024ApJ, de_Graaff2024A&A, Xu2024ApJ, Kohandel2024A&A, Scholtz2025MNRAS},
and perform one-dimensional calculations on a logarithmic radial grid spanning $\rmin = 1~\pc$ to $\rmax = 10^4~\pc$ with 50 bins.
At each radius, we calculate the time evolution of surface mass densities of gas and stars, denoted as $\sgmg$ and $\sgms$, respectively.

We note that our model does not include radial gas advection within galactic disks.
\citetalias{Toyouchi2025MNRAS} investigated the possible impact of this process through post-processing analysis and found that radial advection is likely negligible at $z \gtrsim 7$, where the gas density structure of disks is primarily determined by intense cosmological gas inflow associated with rapid halo mass assembly.

The basic equations we solve are written as follows:
\begin{eqnarray}
\frac{\partial \sgmg}{\partial t} = - \sgmsf + \sgmin - \sgmout + \sgmret \ ,
\label{eq:cons_g}
\end{eqnarray}
\begin{eqnarray}
\frac{\partial \sgms}{\partial t} = \sgmsf - \sgmret \ ,
\label{eq:cons_s}
\end{eqnarray}
where $\sgmsf$, $\sgmin$, and $\sgmout$ are the surface densities of star formation rates, gas inflow rates, and gas outflow rates, respectively.
$\sgmret$ represents the mass return rate due to stellar evolution, which is given by
\begin{eqnarray}
\sgmret = \int^{\mmax}_{m_\ast(t = t_{\rm l})} \mret(m_\ast, Z(t')) \, \sgmsf(t') \, \phi(m_\ast, Z(t')) \frac{{\rm d}m_\ast}{\msun} \ ,
\label{eq:sgmret}
\end{eqnarray}
where $m$ denotes the initial mass of progenitor stars in unit of $\msun$, 
$\phi$ is the mass-normalized IMF, given by $\int^{\mmax}_{\mmin}m_\ast\phi(m_\ast){\rm d}m_\ast = 1~\msun$ ($\mmin = 0.04~\msun$ and $\mmax = 150~\msun$),
$\mret$ is the return mass from a star \citep{Nomoto2013ARA&A},
$t_{\rm l}(m_\ast)$ is the main-sequence stellar lifetime \citep{Schaller1992A&AS}, and we define $t' \equiv t - t_{\rm l}(m_\ast)$.
Here, we adopt the metallicity-dependent IMF proposed by \citet{Chon2024MNRAS}, which transitions from a highly top-heavy IMF (approximately $\phi \propto m_\ast^{-1}$) at $Z < 10^{-3}~\zsun$ to a Salpeter-like IMF ($\phi \propto m_\ast^{-2.35}$) at $Z \sim \zsun$.

In addition, we compute the time evolution of heavy-element abundances.
The governing equation is
\begin{eqnarray}
\frac{\partial (Z_i \sgmg)}{\partial t} = - Z_i \sgmsf + Z_{{\rm in},i} \sgmin - Z_i \sgmout + \sgmzr \ ,
\label{eq:cons_z}
\end{eqnarray}
where $Z_i$ is the mass ratio of the $i$-th element to gas, and we set $Z_{{\rm in}, i} = 0$, assuming metal-free gas inflow for simplicity.
Note that $Z_i$ is regarded as the total metal mass ratio both in gas and dust, so that Eq.~(\ref{eq:cons_z}) does not include the depletion term due to metal accretion on dust grains (see \S~\ref{sec:dest_growth}).
$\sgmzr$ represents the return rate of the $i$-th element via stellar evolution, which is given by
\begin{equation}
\sgmzr = \int^{\mmax}_{m_\ast(t = t_{\rm l})} m_{{\rm ret},i}(m_\ast, Z(t')) \, \sgmsf(t') \, \phi(m_\ast, Z(t')) \frac{{\rm d}m_\ast}{\msun} + \dot{\Sigma}_{{\rm Ia},i} \ 
\label{eq:sgmzr}
\end{equation} 
where $m_{{\rm ret},i}$ is the mass of $i$-th element released by stellar winds and/or SNe II \citep{Nomoto2013ARA&A}.
$\dot{\Sigma}_{{\rm Ia},i}$ represents the return rate of the $i$-th element from SNe Ia, and we modeled it as,
\begin{eqnarray}
\dot{\Sigma}_{{\rm Ia},i} = \int^{t}_{0} m_{{\rm Ia}, i} DTD(t-t') \sgmsf(t') {\rm d}t' \ ,
\label{eq:sgmIa}
\end{eqnarray}
where $m_{{\rm Ia}, i}$ is the mass released by a single SN Ia \citep{Iwamoto1999ApJS}. 
$DTD(t_{\rm delay})$ represents the delay-time distribution for SNe Ia to occur after the star formation, 
and we model it based on the observational estimate by \citet{Maoz2010ApJ}, as follows:
\begin{eqnarray}
DTD(t_{\rm delay}) = 
\begin{cases}
10^{-6} \ (t_{\rm delay}/{\rm Myr})^{-1.1} & (t_{\rm delay} \ge t_{\rm delay,min}) \\
0 & (\text{otherwise})
\end{cases}
\ ,
\label{eq:DTD}
\end{eqnarray}
where we set the minimum delay time to be $t_{\rm delay,min} = 100~\Myr$ \citep{Totani2008PASJ, Maoz2010ApJ}.

Eqs.~(\ref{eq:cons_g}-\ref{eq:cons_z}) can be solved by specifying functional forms of $\sgmsf$, $\sgmin$, and $\sgmout$.
In \citetalias{Toyouchi2025MNRAS}, we modeled $\sgmsf$ with the Kennicutt-Schmidt (KS) law \citep{Schmidt1959ApJ, Kennicutt1998ApJ}, introducing a boost factor of $\mathcal{F}_{\rm b} \geq 1$ defined as the excess relative to the star formation rate expected from the KS law.
Interestingly, we found that the galaxy evolution is almost independent of $\mathcal{F}_{\rm b}$ owing to the self-regulation mechanism: high (low) values of $\mathcal{F}_{\rm b}$ lead to rapid (slow) gas consumption by star formation and decrease (increase) surface gas densities in the galactic disk, which eventually hardly change the SFRs.
We adopt $\mathcal{F}_{\rm b} = 1$ throughout this study.

For the functional forms of $\sgmin$ and $\sgmout$, we refer the readers to \S~2.2 of \citetalias{Toyouchi2025MNRAS} for the detailed introduction.
A notable finding from \citetalias{Toyouchi2025MNRAS} is that star formation efficiencies (SFEs) in galaxies are tightly connected to $\sgmin$ and $\sgmout$ as the balance of these two terms controls the gas mass available for star formation.
Consequently, we reasonably modeled $\sgmin$ and $\sgmout$ to reproduce the observed dependence of SFEs on host dark matter halo masses (see \S~4 of \citetalias{Toyouchi2025MNRAS}).

Finally, we model the mass growth rate of dark matter halos with a fitting formula derived from the cosmological N-body simulation by \citet{Fakhouri2010MNRAS}:
\begin{eqnarray}
\dot{M}_{\rm h} = 46.1~\msun~\yr^{-1}~\left( \frac{M_{\rm h}}{10^{12}~\msun} \right)^{1.1} \nonumber \\
\times (1+1.11~z) \sqrt{\Omega_{\rm m}(1+z)^3+\Omega_\Lambda} \ .
\label{eq:mdoth}
\end{eqnarray}
We evolve 21 dark matter halos under this formula.
The initial mass of $i$-th lightest halo is set as $M_{\rm h, i} = M_{\rm h,1} \times 10^{\delta_i}$ with $\delta_{i} = (i-1)/20 \times {\rm log}(M_{\rm h,21}/M_{\rm h,1})$, where 
$M_{\rm h,1} = 5.8 \times 10^{7}~\msun$ and $M_{\rm h,21} = 5.2 \times 10^{9}~\msun$ correspond to the initial masses of the 1st (lightest) and 21st (heaviest) halos.
With this setup, the 1st, 5th, 9th, 13th, 17th, and 21st halos finally grow to $M_{\rm h} = 10^{10},\,10^{11},\,10^{12},\,10^{13}$ and $10^{14}~\msun$ at $z = 5$, respectively.
Incorporating these diverse halo mass growth histories allows us to examine the galaxy evolution as a function of redshift and their host halo masses.
In the following, we denote the dark matter halo mass at $z = 5$ as $M_{\rm h5} \equiv M_{\rm h}(z = 5)$ for convenience.


\begin{table*}
\centering
\begin{threeparttable}

\caption{Summary of grain properties.}
\label{table:dust}

\begin{tabular}{lccccccc}
\hline
Species &
$X^{a}$ &
$s^{b}$ &
${f_X}^{c}$ &
${v_{\rm shat}}^{d}$ &
${Q_{\rm D}}^{e}$ &
${\gamma}^{f}$ &
${E}^{g}$ \\
 &
 &
(g\,cm$^{-3}$) &
 &
(km\,s$^{-1}$) &
(cm$^{2}$\,s$^{-2}$) &
(erg\,cm$^{-2}$) &
(dyn\,cm$^{-2}$) \\
\hline
Silicate & Si & 3.3 & 0.166 & 2.7 & $4.3\times10^{10}$ & 25 & $5.4\times10^{11}$ \\
Graphite & C  & 2.2 & 1.0   & 1.2 & $8.9\times10^{9}$  & 12 & $3.4\times10^{10}$ \\
\hline
\end{tabular}

\begin{tablenotes}
\footnotesize
\item
$a$: Key species used to calculate grain growth by metal accretion.
$b$: Material density of dust grains.
$c$: Mass fraction of the key species in dust grains.
$d$: Threshold velocity for shattering.
$e$: Critical impact energy.
$f$: Surface energy per unit area.
$g$: Effective Young's modulus.
\end{tablenotes}

\end{threeparttable}
\end{table*}


\subsection{Evolution of grain size distribution}
\label{sec:gsd_evol}

In this section, we describe the model for the evolution of the grain-size distribution, denoted by $n(a,t)$, where $a$ is the grain radius.
For simplicity, we omit the explicit radial dependence of variables introduced below, although all quantities fundamentally depend on galactocentric radius $R$.
For example, the full expression for the grain-size distribution should formally be written as $n(a,R,t)$.

We adopt basically the same model for the evolution of the grain size
distribution as in \citet{Hirashita2019MNRAS} (hereafter H19).
We consider the following dust evolution processes:
(i) dust production by stellar sources,
(ii) dust destruction by SN shocks,
(iii) dust growth by metal accretion,
(iv) dust disruption by shattering, 
and (v) dust growth by coagulation.
Dust grains are assumed to be dynamically coupled with the gas,
which is a valid approximation on the spatial scales considered in
this study \citep{McKinnon2018MNRAS}.
Below we outline the model implemented in our galaxy evolution model;
we refer the readers to \citetalias{Hirashita2019MNRAS} for a complete description and detailed equations.

In this paper, we consider silicate and graphite as representative grain species in the ISM.
To avoid complexity associated with mixed-material interactions, we assume that collisions occur only between the same grain species.
For clarity in the following discussion, we adopt the same notation for all physical quantities and variables for both silicate and graphite,
while separately following the evolution of each grain type by assuming different sets of physical parameters.
Table~\ref{table:dust} summarizes the physical parameters adopted for silicate and graphite, following the values compiled by \citet{Hirashita2009MNRAS}. 

We assume that dust grains are spherical and compact, such that the grain mass $m$ is related to the grain radius $a$ as $m = (4\pi/3) a^3 s $, where $s$ is the material density of dust. 
We adopt $s = 3.3~(2.2)~{\rm g\,cm^{-3}}$ for silicate (graphite).
The grain mass distribution, $\rho_{\rm d}(m,t)$, is then related to the grain size distribution
in radius space, $n(a,t)$, by
\begin{eqnarray}
\rho_{\rm d}(m,t)\,{\rm d}m =
\frac{4}{3}\pi a^3 s\, n(a,t)\,{\rm d}a \ .
\label{eq:rhod_na}
\end{eqnarray}
The total dust mass density is given by
\begin{eqnarray}
\rho_{\rm d,tot}(t) =
\int_0^\infty \rho_{\rm d}(m,t)\,{\rm d}m \ ,
\label{eq:rhod_tot}
\end{eqnarray}
and the dust-to-gas mass ratio is defined as
\begin{eqnarray}
D_{\rm tot}(t) \equiv
\frac{\rho_{\rm d,tot}(t)}{\rho_{\rm g}(t)} ,
\label{eq:dtot}
\end{eqnarray}
where $\rho_{\rm g}$ is the gas mass density (see \S~\ref{sec:two_phase} for the detailed introduction).
We compute $\rho_{\rm d}$ and $D_{\rm tot}$ separately for silicate and graphite, and combine the two components when discussing the overall dust properties.

The time evolution of the grain mass distribution is described by
\begin{align}
\frac{\partial \rho_{\rm d}(m,t)}{\partial t}
&=
\left.\frac{\partial \rho_{\rm d}}{\partial t}\right|_{\rm star}
+
\left.\frac{\partial \rho_{\rm d}}{\partial t}\right|_{\rm sput}
+
\left.\frac{\partial \rho_{\rm d}}{\partial t}\right|_{\rm acc}
\nonumber\\
&\quad
+
\left.\frac{\partial \rho_{\rm d}}{\partial t}\right|_{\rm shat}
+
\left.\frac{\partial \rho_{\rm d}}{\partial t}\right|_{\rm coag}
+
\rho_{\rm d}(m,t)\frac{\rm d\ln\rho_{\rm g}}{{\rm d}t} .
\label{eq:master_eq}
\end{align}
The terms with subscripts ``star'', ``sput'', ``acc'', ``shat'', and
``coag'' represent changes due to stellar dust production,
sputtering destruction, accretion, shattering, and coagulation,
respectively. The last term accounts for variations in the background gas density driven by star formation, gas inflows and outflows, as described by Eq.~(\ref{eq:cons_g}).
In practice, we solve discretized forms of this equation using 32 logarithmically spaced grid points in the grain radius range from $a_{\rm min} = 3\times10^{-4}~\mu$m and $a_{\rm max} = 10~\mu{\rm m}$.

\subsubsection{Stellar dust production}

We focus on type-II SNe as the source of stellar dust production in high-$z$ galaxies as
AGB stars are likely negligible at $z > 5$ due to their long stellar lifetimes (e.g., $\tau \sim 1~\Gyr$ for $m_\ast = 2~\msun$) and lower dust production rates \citep[e.g.,][]{Mancini2015MNRAS, Schneider2024A&ARv}.
We describe the dust production rate by SNe II as,
\begin{eqnarray}
\left.\frac{\partial \rho_{\rm d}(m,t)}{\partial t}\right|_{\rm star}
=
y_{\rm d}\,\dot{n}_{\rm II}\, m(a) \psi(a) \frac{{\rm d}a}{{\rm d}m} \ ,
\label{eq:stellar}
\end{eqnarray}
where $y_{\rm d}$ is the dust mass produced by a single SN II, and 
$\dot{n}_{\rm II}$ is the SN II occurrence rate per unit volume and time, which can be estimated from our galaxy model (see \S~\ref{sec:two_phase} for details).
Here, $\psi(a)$ is the normalized size distribution of produced dust grains,
and is assumed to follow a log-normal form with a mean grain radius $a_0$ and a standard deviation $\sigma$,
\begin{eqnarray}
\psi(a) =
C_\psi \frac{1}{a}
\exp\!\left[
-\frac{\{\ln(a/a_0)\}^2}{2\sigma^2}
\right],
\label{eq:lognormal}
\end{eqnarray}
where $C_\psi$ is a normalization constant.

We note that SN dust yields, $y_{\rm d}$, remain highly uncertain.
While some studies have suggested $y_{\rm d} \sim 1~\msun$ \citep[e.g.,][]{Todini2001MNRAS, Bianchi2007MNRAS}, 
significantly smaller yields may help explain the extremely low dust-to-stellar mass ratios of $\lesssim 10^{-4}$ inferred for galaxies at $z > 10$ \citep[e.g.,][]{Algera2023MNRAS, Bakx2025MNRAS, Bakx2026MNRAS}.
One possible explanation for such reduced dust yields is efficient dust destruction by reverse shocks in dense ISM environments \citep[e.g.,][]{Nozawa2006ApJ, Nozawa2007ApJ, Leniewska2019A&A, Slavin2020ApJ, Scheffler2026A&A}.
Since high-redshift galaxies are generally expected to have denser ISM conditions than local galaxies,\footnote{This argument may not always hold, because OB-stars can reduce the surrounding gas density through photoionization and stellar winds before SN explosions \citep{Martinez2019ApJ}.} newly formed dust grains may undergo stronger reverse-shock destruction.
Indeed, \citet{Otaki2026A&A} predicted dust yields of $y_{\rm d} < 10^{-3}~\msun$ under dense ISM conditions, highlighting the importance of reverse-shock destruction in regulating dust production by SNe II.

Similarly, the typical size of SN dust is also uncertain, with estimates ranging from $a_0 = 0.01$-$0.1~\mu {\rm m}$ \citep[e.g.,][]{Nozawa2007ApJ, Asano2013EP&S, Hirashita2019MNRAS, Otaki2026A&A}.
To explore these uncertainties, we vary $y_{\rm d} = 10^{-6}$-$10^{-2}~\msun$ and $a_0 = 0.01$-$0.1~\mu {\rm m}$, adopting fiducial values of $y_{\rm d} = 10^{-4}~\msun$ and $a_0 = 0.01~\mu{\rm m}$.
Throughout this paper, we fix $\sigma = 0.5$ for simplicity,
although in reality it may vary with $y_{\rm d}$ and $a_0$.
We further comment on this point in more details in \S~\ref{sec:para}.

\subsubsection{Dust destruction and growth}\label{sec:dest_growth}

Dust destruction by sputtering in SN shocks and dust growth by
accretion are described by the same form of equation,
\begin{eqnarray}
\left.\frac{\partial \rho_{\rm d}(m,t)}{\partial t}\right|_{\rm sput/acc}
=
-\frac{\partial}{\partial m}
\left[m\,\dot{\rho}_{\rm d}(m,t)\right]
+
\frac{\dot{m}}{m}\rho_{\rm d}(m,t),
\label{eq:sput_acc}
\end{eqnarray}
where $\dot{m} \equiv dm/dt$ is the mass change rate of an individual grain.

For dust destruction, we describe the mass change rate as
\begin{eqnarray}
\dot{m} =
- \frac{m}{\tau_{\rm dest}(m)} ,
\label{eq:mdot_dest}
\end{eqnarray}
where the mass-dependent dust destruction timescale $\tau_{\rm dest}$ is estimated as
\begin{eqnarray}
\tau_{\rm dest}(m) =
\frac{\rho_{\rm g}}
{\epsilon_{\rm dest}(m)\,M_s\,\dot{n}_{\rm II}}.
\label{eq:taudest}
\end{eqnarray}
$M_s = 6800~\msun$ is the gas mass swept up by a single SN II \citep[][]{McKee1989ApJ, Nozawa2006ApJ}, and
$\epsilon_{\rm dest}(m)$ is the grain-size-dependent destruction efficiency. 
For $\epsilon_{\rm dest}(m)$, we adopt the empirical formula proposed by \citetalias{Hirashita2019MNRAS},
\begin{eqnarray}
\epsilon_{\rm dest}(a) =
1 - \exp\!\left[
-0.1\left(\frac{a}{0.1~\mu{\rm m}}\right)^{-1}
\right],
\label{eq:edest}
\end{eqnarray}
which yields $\epsilon_{\rm dest} \sim 1$ for small grains with $a \ll 0.1~\mu {\rm m}$, whereas $\epsilon_{\rm dest} \propto a^{-1}$ for large ones with $a > 0.1~\mu {\rm m}$.

In this study, we simply evaluate the dust growth rate by considering the accretion of the key species denoted as $X$, which is Si for silicate and C for graphite \citep[][]{Hirashita2011MNRAS}.
The mass fraction of the key species in dust is denoted by $f_X$, with $f_X = 0.166$ for silicate 
(i.e., Si constitutes a mass fraction of 0.166 in silicate), while $f_X = 1$ for graphite (i.e., graphite is composed of only C).
Under this assumption, the mass growth rate due to metal accretion is written as
\begin{eqnarray}
\dot{m} = \left ( 1-\frac{f_X D_{\rm tot}}{Z_X} \right ) \frac{m}{\tau_{\rm acc}(m)} ,
\label{eq:mdot_acc}
\end{eqnarray}
where $\tau_{\rm acc}$ is the accretion timescale.
The factor in parentheses accounts for the depletion of the key species in the ISM caused by metal accretion. 
Following \citet{Hirashita2012MNRAS}, we express the accretion timescale as
\begin{eqnarray}
\tau_{\rm acc}(m) &=& 
\frac{1}{3} \frac{a f_X s}{Z_{X}\, \rho_{\rm g}\, S} \left( \frac{2\pi m_X}{k_{\rm B} T_{\rm g}} \right)^{1/2} \nonumber \\ 
&=&
\frac{1}{3} \tau_{0,{\rm acc}}
\left(\frac{a}{0.1~\mu{\rm m}}\right)
\left(\frac{Z_X}{Z_{X, \odot}}\right)^{-1} \nonumber \\
&& \times  
\left(\frac{n_{\rm g}}{10^3~{\rm cm^{-3}}}\right)^{-1}
\left(\frac{T_{\rm g}}{10~{\rm K}}\right)^{-1/2}
\left(\frac{S}{0.3}\right)^{-1}
\ ,
\label{eq:tauacc}
\end{eqnarray}
where $m_X$ is the atomic mass of $X$, $S$ is the sticking probability for
accretion, and $k_{\rm B}$ is the Boltzmann constant. 
The gas number density and the gas temperature are denoted as $n_{\rm g}$ and $T_{\rm g}$, respectively (see \S~\ref{sec:two_phase} for detailes).
The solar abundance of the key species $Z_{X, \odot}$ is  taken from \citet{Asplund2005ASPC}.
The normalization constant is evaluated as $\tau_{0,{\rm acc}} = 1.61 \times 10^8~\yr$ for silicate and $0.993 \times 10^8~\yr$ for graphite.
Throughout this paper, we adopt $S = 0.3$ for both silicate and graphite \citep[][]{Leitch-Devlin1985MNRAS, Grassi2011A&A}.

Note that the accretion timescale becomes shorter with decreasing grain size, reflecting the fact that smaller grains have a effectively larger surface-area-to-volume ratio.
As a consequence, smaller grains grow more efficiently in the ISM than larger grains.

\subsubsection{Shattering}

Shattering is driven by grain--grain collisions that result in the fragmentation of grains.
The time evolution of the grain mass distribution due to shattering is written as
\begin{align}
\left.\frac{\partial \rho_{\rm d}(m,t)}{\partial t}\right|_{\rm shat}
&=&
- m\rho_{\rm d}(m,t)
\int_0^\infty
\alpha_{\rm coag}(m_1,m)\rho_{\rm d}(m_1,t)\,{\rm d}m_1
\nonumber\\
&& + 
\int_0^\infty\!\!\int_0^\infty
\alpha_{\rm shat}(m_1,m_2)
\rho_{\rm d}(m_1,t)\rho_{\rm d}(m_2,t) \nonumber \\
&& \times \, \mu_{\rm shat}(m;m_1,m_2)\,{\rm d}m_1 {\rm d}m_2 ,
\label{eq:shat}
\end{align}
where $\mu_{\rm shat}(m;m_1,m_2)$ describes the mass distribution of fragments
produced by shattering, and $\alpha_{\rm shat}(m_1,m_2)$ is the collision kernel.
In Eq.~(\ref{eq:shat}), the first term represents the loss of grains of mass $m$ due to shattering, 
and the second term accounts for the production of grains of mass $m$
as fragments generated in collisions between grains of masses $m_1$ and $m_2$.

Here, we define the collision kernel as
\begin{eqnarray}
\alpha_{\rm shat}(m_1,m_2) =
\begin{cases}
\frac{\sigma_{1,2}v_{1,2}}{m_1m_2} & (v_{1,2} \ge v_{\rm shat}) \\
0 & (\text{otherwise})
\end{cases}
\ ,
\label{eq:alpha_shat}
\end{eqnarray}
where the geometric cross section is $\sigma_{1,2}=\pi(a_1+a_2)^2$, 
the relative velocity is $v_{1,2}$,
and $v_{\rm shat}$ is the velocity threshold, above which shattering occurs.
We adopt $v_{\rm shat} = 2.7~(1.2)~{\rm km\,s^{-1}}$ for silicate (graphite).

The velocity of a grain in turbulent media is approximated as \citep{Ormel2009A&A},
\begin{eqnarray}
v_{\rm gr}(a) &=&
1.1 \mathcal{M}^{3/2}
\left(\frac{a}{0.1~\mu{\rm m}}\right)^{1/2}
\left(\frac{T_{\rm g}}{10^4~{\rm K}}\right)^{1/4} \nonumber \\ 
&& \times 
\left(\frac{n_{\rm g}}{1~{\rm cm^{-3}}}\right)^{-1/4}
\left(\frac{s}{3.5~{\rm g\,cm^{-3}}}\right)^{1/2}
\,{\rm km\,s^{-1}},
\label{eq:vgr}
\end{eqnarray}
where $\mathcal{M}$ is the Mach number associated with the largest turbulent eddies, whose characteristic scale is expected to be comparable to the local Jeans length. 
The adopted values of $\mathcal{M}$ are presented in \S~\ref{sec:two_phase}.

For a collision between two grains with $v_1 = v_{\rm gr}(a_1)$ and $v_2 = v_{\rm gr}(a_2)$,
the relative velocity is estimated as,
\begin{eqnarray}
v_{1,2} = \sqrt{v_1^2+v_2^2-2 v_1 v_2 \mu_{1,2}} \, 
\label{eq:vrel}
\end{eqnarray}
where $\mu \equiv {\rm cos}\theta$ represents the collision angle $\theta$ between the two grains. 
In each evaluation of $\alpha$, we randomly sample $\mu_{1,2}$ from the interval $[-1:1]$.

We suppose that a collision between two grains with $m_1$ and $m_2$ yields the total ejected mass from $m_1$ as
\begin{eqnarray}
m_{\rm ej} = \frac{x}{1+x} m_1 \ ,
\label{eq:mej}
\end{eqnarray}
where $x \equiv E_{\rm imp}/(m_1\,Q_{\rm D})$, $E_{\rm imp} = \frac{1}{2} (m_1 m_2)/(m_1+m_2) v_{1,2}^2 $ is the impact parameter, and $Q_{\rm D}$ is the critical impact energy, above which the disrupted mass exceeds $m_1/2$.
Eq.~(\ref{eq:mej}) yields $m_{\rm ej} \sim E_{\rm imp}/Q_{\rm D}$ in the weak collision limit ($x \ll 1$),
whereas $m_{\rm ej} \sim m_1$ in the strong collision limit ($x \gg 1$).
We adopt $Q_{\rm D} = 4.3 \times 10^{10}~{\rm cm^2\,s^{-2}}$ for silicate and $8.9 \times 10^{9}~{\rm cm^2\,s^{-2}}$ for graphite \citep{Jones1996ApJ}.

Then, we assume that grain sizes of the shattered fragments follows a power-law distribution  with an index of $\alpha_{\rm f}$, corresponding to the index of mass distribution of $(\alpha_{\rm f}+1)/3$,
where the maximum and minimum grain masses of the fragments are $m_{\rm f,max} = 0.02 m_{\rm ej}$ and $m_{\rm f,min} = 10^{-6}\,m_{\rm f,max}$, respectively \citep[][]{Guillet2011A&A}.
Under this assumption, the fragment mass distribution including the remnant of mass $m_1-m_{\rm ej}$ is written as
\begin{align}
\mu_{\rm shat}(m; m_1, m_2)
&=
\frac{(4+\alpha_{\rm f})\,m_{\rm ej}\,m^{(\alpha_{\rm f}+1)/3}}
{
3 \left [ m_{\rm f,max}^{\,(4+\alpha_{\rm f})/3} - m_{\rm f,min}^{\,(4+\alpha_{\rm f})/3} \right ]
}
\,
\Theta(m; m_{\rm f,min}, m_{\rm f,max})
\nonumber\\
&\quad
+
\left(m_1 - m_{\rm ej}\right)\,
\delta\!\left(m - m_1 + m_{\rm ej}\right) \ ,
\label{eq:mu_shat}
\end{align}
where $\Theta(m; m_{\rm f,min}, m_{\rm f,max}) = 1$ if $m_{\rm f,min} < m < m_{\rm f,max}$,
and $0$ otherwise, $\delta$ is Dirac's delta function.
In this study, we adopt $\alpha_{\rm f} = -3.3$ \citep[][]{Jones1996ApJ}.

\subsubsection{Coagulation}

Coagulation occurs when two dust grains collide at sufficiently low relative velocities to stick to each other.
Analogously to shattering, the time evolution of the grain mass distribution due to coagulation is written as
\begin{align}
\left.\frac{\partial \rho_{\rm d}(m,t)}{\partial t}\right|_{\rm coag}
&=&
- m\rho_{\rm d}(m,t)
\int_0^\infty
\alpha_{\rm coag}(m_1,m)\rho_{\rm d}(m_1,t)\,{\rm d}m_1
\nonumber\\
&& + 
\int_0^\infty\!\!\int_0^\infty
\alpha_{\rm coag}(m_1,m_2)
\rho_{\rm d}(m_1,t)\rho_{\rm d}(m_2,t) \nonumber \\
&& \times \, m_{1} \delta(m-m_1-m_2)\,{\rm d}m_1 {\rm d}m_2 ,
\label{eq:coag}
\end{align}
and the collision kernel for coagulation is given by
\begin{eqnarray}
\alpha_{\rm coag}(m_1,m_2) =
\begin{cases}
\frac{\sigma_{1,2}v_{1,2}}{m_1m_2} & (v_{1,2} \le v_{\rm coag}^{1,2}) \\
0 & (\text{otherwise})
\end{cases}
\ ,
\label{eq:alpha_coag}
\end{eqnarray}
where $v_{\rm coag}^{1,2}$ is the velocity threshold, below which coagulation occurs.
The coagulation threshold velocity is given by \citet{Chokshi1993ApJ} as
\begin{eqnarray}
v_{\rm coag}^{1,2} = 2.14 \, F_{\rm stick} \,
\left [ \frac{a_1^3+a_2^3}{(a_1+a_2)^3} \right ]^{1/2} 
\frac{\gamma^{5/6}}{E^{1/3}R_{1,2}^{5/6}s^{1/2}}~{\rm cm\,s^{-1}} \ ,
\label{eq:vcoag}
\end{eqnarray}
where we adopt a sticking factor of $F_{\rm stick} = 10$ \citep[][]{Blum2000SSRv, Yan2004ApJ}.
Here, $\gamma$ is the surface energy per unit area,
$R_{1,2} \equiv a_1\,a_2/(a_1+a_2)$ is the reduced radius of the grains,
and $E$ corresponds to the effective Young's modulus that characterizes the elastic response of the grain--grain contact.
The values of $\gamma$ and $E$ are taken from \citet{Chokshi1993ApJ} and are listed in Table~\ref{table:dust}.
The relative velocity $v_{1,2}$ used to evaluate coagulation is calculated in the same manner as for shattering, using Eqs.~(\ref{eq:vgr}) and (\ref{eq:vrel}).

\begin{table*}
\centering
\caption{Key parameters in the dust evolution model.}
\label{table:param}
\renewcommand{\arraystretch}{1.3}
\begin{tabular}{cccl}
\hline
Parameter & Fiducial value & Explored range & Physical meaning \\
\hline
$\mhfive$ & $10^{13}~\msun$ & $10^{10}$--$10^{14}~\msun$ & Mass of the host dark matter halo at $z = 5$ \\
$\fcm$ & $0.9$ & $0.1$--$1$ & Mass fraction of the CMG in the ISM \\
$\yd$ & $10^{-4}~\msun$ & $10^{-6}$--$10^{-2}~\msun$ & Dust yield per Type~II supernova \\
$a_0$ & $0.01~\mu{\rm m}$ & $0.01$--$0.1~\mu{\rm m}$ & Mean grain size of dust formed by Type~II supernovae \\
\hline
$z_{\rm s}$ & $20$ & $-$ & Start redshift of the model calculations \\
$z_{\rm e}$ & $5$ & $-$ & End redshift of the model calculations \\
$R_{\rm min}$ & $1~\pc$ & $-$ & Minimum galactocentric radius \\
$R_{\rm max}$ & $10^4~\pc$ & $-$ & Maximum galactocentric radius \\
$\Tc$ & ${\rm max}(50~{\rm K},\, T_{\rm CMB})$ & $-$ & CMG temperature \\
$\Tw$ & $10^4~{\rm K}$ & $-$ & WNG temperature \\
$\fcv$ & $0.1$ & $-$ & Volume fraction of the CMG in the ISM \\
$\mathcal{M}_{\rm c}$ & $1$ & $-$ & Mach number in the CMG \\
$\mathcal{M}_{\rm w}$ & $3$ & $-$ & Mach number in the WNG \\
$H$ & $0.5~R$ & $-$ & Disk scale height \\
$\rc$ & $0.1~H$ & $-$ & Radius of CMG clouds \\
$\tau^{\rm vy}_\ast$ & $10~\Myr$ & $-$ & Maximum age of very young stars associated with CMG clouds \\
\hline
\end{tabular}
\renewcommand{\arraystretch}{1.0}
\end{table*}

\subsection{Two-phase ISM model}\label{sec:two_phase}

As shown in the previous section, dust growth and destruction depend strongly on the ISM gas density $\rhog$ and temperature $\Tg$.
Since galaxies generally host multiphase gas with a wide range of densities and temperatures, the nature of dust evolution is expected to vary among different regions.
To incorporate such multiphase effects in a phenomenological manner, we assume
that the ISM consists of two components: cold molecular gas (CMG) and warm neutral gas (WNG).
For simplicity, we ignore the hot ionized phase suggested by some recent observations of high-redshift galaxies \citep{Topping2025ApJ, Usui2025ApJ, Harikane2025ApJ}.

We consider the dust mass density distributions in the CMG and the WNG, denoted as $\rhodc(m,\,t)$ and $\rhodw(m,\,t)$, respectively.
Since the CMG forms from the WNG through radiative cooling, $\rhodc(m,\,t)$ in a newly formed CMG cloud is initially identical to $\rhodw(m,\,t)$.
Subsequently, $\rhodc(m,\,t)$ can gradually deviate from $\rhodw(m,\,t)$
owing to efficient dust growth in the CMG.
When the cloud is eventually dispersed by stellar feedback and returns to the WNG phase, the evolved grain population in the CMG is mixed back into the diffuse ISM \footnote{\citet{Ferrara2016MNRAS} pointed out that, in dense CMG clouds, materials accreted onto dust grains are likely to form volatile mantles that can be immediately photodesorbed once the grains return to the diffuse WNG after cloud dispersal.
This raises the question of whether dust growth occurring within the CMG can effectively contribute to the global dust evolution in galaxies.
}.
The difference between $\rhodc(m,\,t)$ and $\rhodw(m,\,t)$ becomes significant when the metal accretion timescale in the CMG is much shorter than the cloud lifetime, which is typically 1-10 Myr.
This condition is occasionally satisfied in our model, particularly in the inner disk regions within the half-SFR radius of massive galaxies, as discussed in \S~\ref{sec:results}.

In this paper, however, we neglect such phase-to-phase variations in the grain size distributions for simplicity, and leave a more detailed treatment of the stochastic formation and destruction of CMG clouds to future work.
Instead, we consider the ISM-averaged dust mass density distribution, $\rhodav(m,\,t)$, and assume that the shapes of $\rhodc$ and $\rhodw$ are always identical to that of $\rhodav$ while the normalization can deviate reflecting the density contrast between the two phases.
We, then, evaluate the time derivatives of $\rhodc$ and $\rhodw$ separately and combine them in a mass-weighted manner to update $\rhodav$.
In this way, we phenomenologically incorporate the cumulative effect of enhanced dust growth in CMG clouds into the long-term evolution of dust in galaxies.

Specifically, we define $\rhodc$ and $\rhodw$ at each timestep such that their normalizations scale with the gas density:
\begin{eqnarray}
\rhodc(m,\,t) = \rhodav(m,\,t) \frac{\rhoc}{\rhoavg} \, ,
\label{eq:rhodc}
\end{eqnarray}
and
\begin{eqnarray}
\rhodw(m,\,t) = \rhodav(m,\,t) \frac{\rhow}{\rhoavg} \, ,
\label{eq:rhodw}
\end{eqnarray}
where $\rhoc$ and $\rhow$ are the gas densities in the CMG and WNG, respectively, and $\rhoavg$ is the mean ISM density.
The evolution of $\rhodav(m,\,t)$ is then written as,
\begin{eqnarray}
\frac{\partial \rhodav(m,\,t)}{\partial t} = 
\fcv \, \frac{\partial \rhodc(m,\,t)}{\partial t} +
(1-\fcv) \, \frac{\partial \rhodw(m,\,t)}{\partial t} \, ,
\label{eq:drhod_dt}
\end{eqnarray}
where $\fcv$ is the volume fraction of the CMG in the ISM, such that the corresponding volume fraction of the WNG is $1-\fcv$.
The time derivatives of $\rhodc$ and $\rhodw$ are evaluated using Eq.~(\ref{eq:master_eq}) once the physical parameters ($\rhog$, $\Tg$, $\dot{n}_{\rm II}$, and $\mathcal{M}$) are specified for each phase.

\subsubsection{Physical properties of the two phases}

We fix the WNG temperature to $\Tw = 8000~{\rm K}$ and define the CMG temperature as $\Tc = {\rm max}(50~{\rm K},\, T_{\rm CMB})$,
where the CMB temperature is $T_{\rm CMB} = T_0 (1+z)$ with $T_0 = 2.7255~{\rm K}$.
We assume supersonic turbulence in the WNG with Mach number $\mathcal{M}_{\rm w}=3$, while the CMG is assumed to be less turbulent with $\mathcal{M}_{\rm c}=1$, following the assumption adopted by \citetalias{Hirashita2019MNRAS}. 
These values are motivated by the study of \citet{Yan2004ApJ}, who investigated grain acceleration by hydrodrag and gyroresonance in magnetohydrodynamic turbulence and calculated the grain velocities achieved in CMG and WNG.

By defining the CMG mass fraction as $\fcm$, as well as the corresponding volume fraction $\fcv$,
the gas densities in the two phases are written as
\begin{eqnarray}
\rhoc = \frac{\fcm}{\fcv} \rhoavg \ ,
\label{eq:rhoc}
\end{eqnarray}
\begin{eqnarray}
\rhow = \frac{1-\fcm}{1-\fcv} \rhoavg \ ,
\label{eq:rhow}
\end{eqnarray}
where the mean ISM density is 
\begin{eqnarray}
\rhoavg = \frac{\Sigma_{\rm g}}{2 H } \ .
\label{eq:rhoavg}
\end{eqnarray}
Here, $H$ is the disk scale height; we adopt $H = 0.5~R$, supposing geometrically thick disks in high-redshift galaxies.
The gas number densities are $\nc = \rhoc/(\muc m_{\rm p})$ and $\nw = \rhow/(\muw m_{\rm p})$,
where $m_{\rm p}$ is the proton mass, and the mean molecular weights are $\muc = 2.3$ and $\muw = 1.2$.

The value of $\fcm$ may vary widely depending on galaxy properties.
Local star-forming galaxies typically have $\fcm \sim 0.1$--$0.5$ \citep{Heiles2003ApJ, Saintonge2011MNRAS, Boselli2014A&A, Murray2018ApJS, Catinella2018MNRAS}.
On the other hand, theoretical models have suggested that the CMG mass fraction increases with redshift and can reach $\fcm \gtrsim 0.9$ at $z \gtrsim 6$
\citep{Obreschkow2009ApJ, Lagos2011MNRAS, Popping2014MNRAS}.
To explore this diversity, we vary $\fcm$ over $0.1$--$1$ and neglect its time evolution and halo-mass dependence.
In addition, the volume fraction $\fcv$ may vary, but to isolate the effect of $\fcm$ we fix $\fcv=0.01$, yielding $\rhoc\sim10$--$100\,\rhow$. 
The density contrast approximately corresponds to the pressure equilibrium between the CMG and WNG phases, i.e., $\rhoc \Tc \sim \rhow \Tw$.

Next, we assume that very young stars with ages of $\tau_\ast \le \tau_\ast^{\rm vy}$ are embedded
in CMG clouds, whereas older stars are uniformly distributed in the WNG.
Under this assumption, the Type~II supernova rates in each phase are evaluated as,
\begin{eqnarray}
\nIIc = \frac{1}{\fcv} \, \int^{m_{\rm II,max}}_{m_\ast^{\rm vy}}
\frac{\dot{\Sigma}_{\rm sf}(t')}{2\,H} \, \phi(m_\ast, Z(t')) \,
\frac{{\rm d}m_\ast}{\rm M_\odot} \ ,
\label{eq:nIIc}
\end{eqnarray}
\begin{eqnarray}
\nIIw = \frac{1}{1-\fcv} \, \int^{m_\ast^{\rm vy}}_{m_{\rm II,min}}
\frac{\dot{\Sigma}_{\rm sf}(t')}{2\,H} \, \phi(m_\ast, Z(t')) \,
\frac{{\rm d}m_\ast}{\rm M_\odot} \ ,
\label{eq:nIIw}
\end{eqnarray}
where $m_{\rm II,min}=10~M_\odot$ and $m_{\rm II,max}=40~M_\odot$.
We define very young stars by $\tau_\ast^{\rm vy}=10~{\rm Myr}$, corresponding
to $m_\ast^{\rm vy} \simeq 15~M_\odot$.

Finally, Eq.~(\ref{eq:master_eq}) is solved separately for the CMG using
$\{\rhodc,\rhoc,\Tc,\mathcal{M}_{\rm c},\nIIc\}$ and for the WNG using
$\{\rhodw,\rhow,\Tw,\mathcal{M}_{\rm w},\nIIw\}$, and the ISM-averaged
distribution $\rhodav(m,t)$ is updated using Eq.~(\ref{eq:drhod_dt}).

\subsubsection{Calculations of dust attenuation}

In this study, we use the grain size distribution calculated with the two-phase ISM model described above to compute dust attenuation in galaxies and directly compare the results with observations.
For a given grain size distribution, the absorption and scattering opacities per unit dust mass, denoted as $\kabs$ and $\ksca$, respectively, are calculated as
\begin{eqnarray}
\kas = \sum_{i} \int^{a_{\rm max}}_{a_{\rm min}} 
\frac{\pi \, a^2 \, n_{i}(a)}{D_{{\rm tot},i} \, \rhog} 
 \, \Qas \, {\rm d}a \ ,
\label{eq:kabs_sca}
\end{eqnarray}
where the subscript $i$ indicates the contributions from graphite and silicate.
$\Qabs$ and $\Qsca$ are the absorption and scattering efficiency factors, respectively, calculated from the Mie theory using the same optical constants as in \citet{Weingartner2001ApJ} for graphite and silicate.

In this paper, we define the effective absorption opacity for radiation propagating through the ISM as
\begin{eqnarray}
\keff = \sqrt{\kabs \, (\kabs + \ksca)} \ ,
\label{eq:kappa}
\end{eqnarray}
which effectively accounts for the increase in photon path length due to scattering \citep{Rybicki1979}.
Note that, as in Eqs.~(\ref{eq:rhodc}) and (\ref{eq:rhodw}), we assume that the overall shape of the grain size distribution and the dust-to-gas mass ratio are identical among the CMG and WNG regions. 
Therefore, the absorption opacity can be represented by $\keff$ throughout the entire region.

We assume that the CMG always exists on the galactic midplane in the form of gas clouds of size $\rc$.
We further assume that the very young stars are associated with these clouds over the same spatial extent,
whereas the older stars are uniformly distributed except the clouds.
Under these assumptions, the escape fractions for radiation from the very young stars and the older ones can be estimated as
\begin{eqnarray}
f_{\lambda,\rm esc}^{\rm vy} 
=
\frac{1-\exp(-\tauc)}{\tauc}
\exp\left(
-\tauw\frac{H-\rc}{H}
\right)
\label{eq:fesc_vy}
\end{eqnarray}
\begin{eqnarray}
f_{\lambda,\rm esc}^{\rm o}
=
\frac{1-\exp(-\tauw)}{\tauw} \, ,
\label{eq:fesc_old}
\end{eqnarray}
where the optical depths in the CMG and WNG are defined as
\begin{eqnarray}
\tauc \equiv \keff D_{\rm tot}\rho_c \rc
\label{eq:tauc}
\end{eqnarray}
\begin{eqnarray}
\tauw \equiv \keff D_{\rm tot}\rho_w H \, .
\label{eq:tauw}
\end{eqnarray}
The value of $\rc$ is an important parameter that determines $\tauc$.
In principle, it should be set by the balance between the cloud self-gravity and pressure, although our model does not explicitly consider this condition.
Instead, we fix $\rc / H = 0.1$ for simplicity, which roughly corresponds to the typical ratio between galactic disk thickness and the size of open clusters \citep[][]{Piskunov2007A&A, Castro-Ginard2020A&A, Tarricq2022A&A}.

Recall that, in this model, all variables depend on galactocentric radius $R$.
Accordingly, the galactic luminosity at a given wavelength is obtained by integrating the local radiation over the entire disk.
The total luminosity can therefore be written as the sum of contributions from very young and older stellar populations:
\begin{eqnarray}
L_\lambda
=
2\pi
\int_{R_{\min}}^{R_{\max}}
\left (f_{\lambda,\rm esc}^{\rm vy} \, I_\lambda^{\rm vy}
+ f_{\lambda,\rm esc}^{\rm o} \, I_\lambda^{\rm o} \right )
R\,{\rm d}R  \, .
\label{eq:Luminosity}
\end{eqnarray}
Here, the surface brightnesses at any radius are
\begin{eqnarray}
I_\lambda^{\rm vy}(t)
=
\int_0^{\tau^{\rm vy}_\ast}
\mathcal{L}_\lambda(t',\,Z(t-t'))
\dot{\Sigma}_{\rm sf}(t-t)
{\rm d}t'
\label{eq:Ivy}
\end{eqnarray}
\begin{eqnarray}
I_\lambda^{\rm o}(t)
=
\int_{\tau^{\rm vy}_\ast}^{t}
\mathcal{L}_\lambda(t',\,Z(t-t'))
\dot{\Sigma}_{\rm sf}(t-t')
{\rm d}t' \, ,
\label{eq:Iold}
\end{eqnarray}
where $\mathcal{L}_\lambda(t',\,Z)$ is the specific luminosity per unit stellar mass and unit wavelength of a stellar population with age $t'$ and metallicity $Z$. We compute it with the public Python package {\it Flexible Stellar Population Synthesis} \citep[FSPS;][]{Conroy2010ascl} based on the metallicity-dependent IMF adopted in this study \citep[][]{Chon2024MNRAS}.

Based on these SED calculation, we investigate the radiative properties of galaxies at $z > 5$.
In this paper, we define the UV luminosity as $L_{\rm UV} = \lambda L_\lambda$ at the reference wavelength $\lambda = 1500~\textrm{\AA}$.
We also characterize the spectral slope in the UV by
\begin{eqnarray}
\beta_{\rm UV}
= \frac{\log (L_{\lambda_2}/L_{\lambda_1})}{\log(\lambda_2/\lambda_1)} \, ,
\label{eq:betauv}
\end{eqnarray}
where we set $\lambda_1 = 1230$~\AA~and $\lambda_2 = 3200$~\AA.
Finally, the IR luminosity of each galaxy is evaluated by assuming that UV radiation ($\lambda = 912$--$4000~\textrm{\AA}$) is absorbed by dust and re-emitted in the IR: 
\begin{eqnarray}
L_{\rm IR}
= \int^{4000}_{912} (L^{\rm int}_\lambda-L_\lambda) \, {\rm d}\lambda \, ,
\label{eq:L_IR}
\end{eqnarray}
where $L^{\rm int}_\lambda$ is the intrinsic luminosity without dust attenuation, evaluated by setting $f_{\lambda,\rm esc}^{\rm vy} = f_{\lambda,\rm esc}^{\rm o} = 1$ in Eq.~(\ref{eq:Luminosity}).

\subsubsection{Extinction curve and Attenuation curve}

In this paper, we explicitly distinguish between the extinction curve and the attenuation curve,
denoted as $A^{\rm ext}_{\lambda}$ and $A^{\rm att}_{\lambda}$, respectively,
both of which are commonly used for comparison with observations.
The extinction curve is directly linked to the grain size distribution and is defined as
\begin{eqnarray}
A^{\rm ext}_{\lambda} = 2.5 \, {\rm log}_{10}(e) \, H \, \sum_{i} \int^{a_{\rm max}}_{a_{\rm min}} 
\pi \, a^2 \, n_{i}(a) \, \Qext \, {\rm d}a \ ,
\label{eq:ext_curve}
\end{eqnarray}
where the extinction efficiency is given by $Q^{\rm ext}_{i} = \Qabs + \Qsca$.

On the other hand, the dust attenuation curve depends not only on 
the grain size distribution but also on the star formation history and the spatial distribution of dust relative to stars.
We evaluate the attenuation curve as
\begin{eqnarray}
A^{\rm att}_{\lambda} = - 2.5 \, {\rm log}_{10} \left ( \frac{L_\lambda}{L^{\rm int}_\lambda}\right ) \ .
\label{eq:att_curve}
\end{eqnarray}

In \S~\ref{sec:size} and \ref{sec:att_curve}, we present the predicted extinction and attenuation curves
and discuss how the evolving grain size distribution affects the radiative properties of high-$z$ galaxies.


\begin{figure}
\centering
\includegraphics[width=0.9\columnwidth]{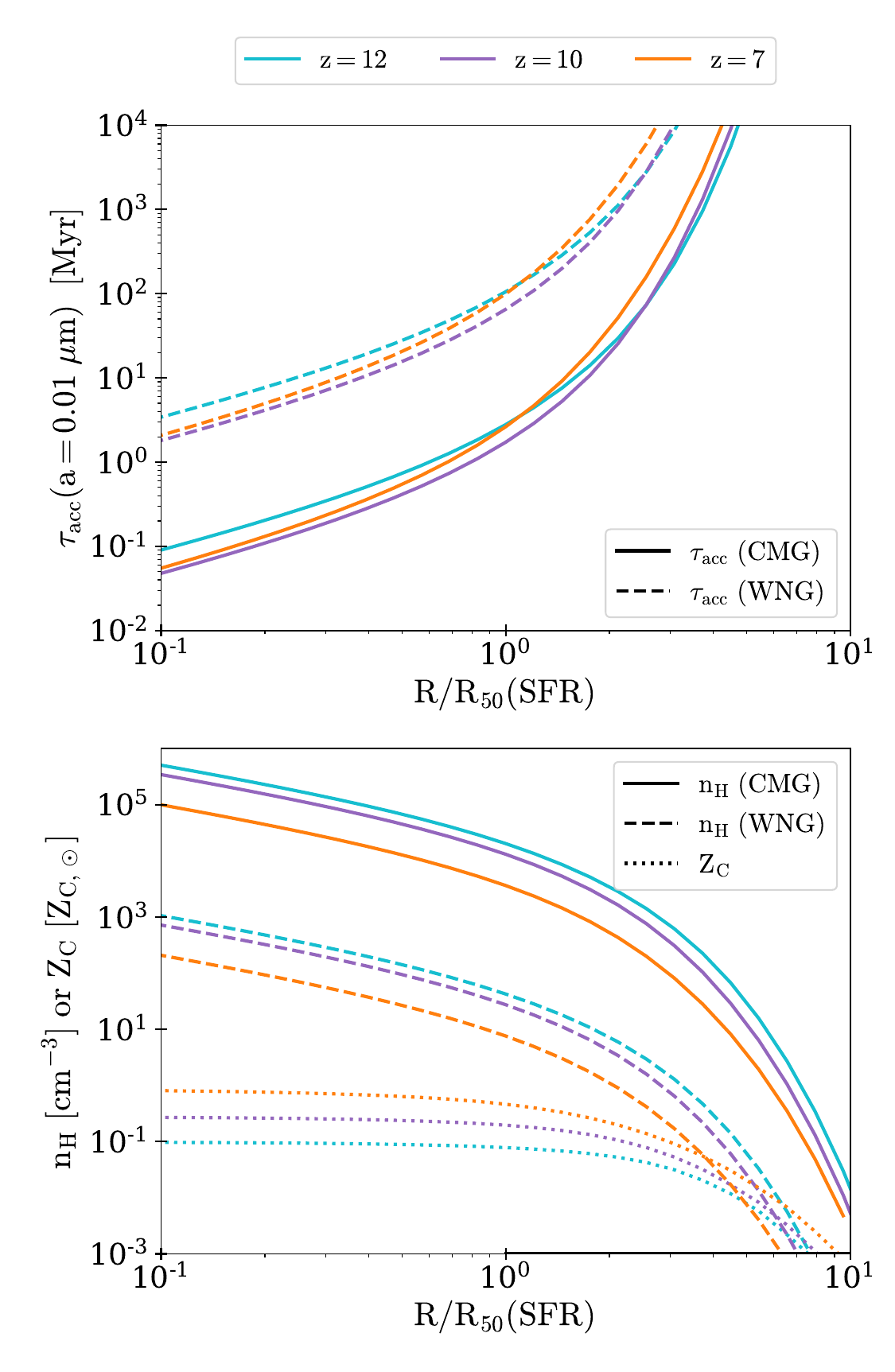}
\caption{
{\it Upper panel}: Radial profiles of the metal accretion timescale on graphite grains with a size of $a = 0.01~\mu$m in the CMG (solid) and WNG (dashed) phases, calculated for a dark matter halo with $\mhfive = 10^{13}~\msun$ under the fiducial parameter set.
The cyan, purple, and orange curves represent the profiles at $z = 12$, 10, and 7, respectively.
The horizontal axis is normalized by the half-SFR radius at each redshift.
{\it Lower panel}: Same as the upper panel, but showing the radial profiles of the gas number density in the CMG (solid) and WNG (dashed), together with the carbon abundance (dotted).
}
\label{fig:RP_nzt}
\end{figure}

\begin{figure*}
\centering
\includegraphics[width=1.8\columnwidth]{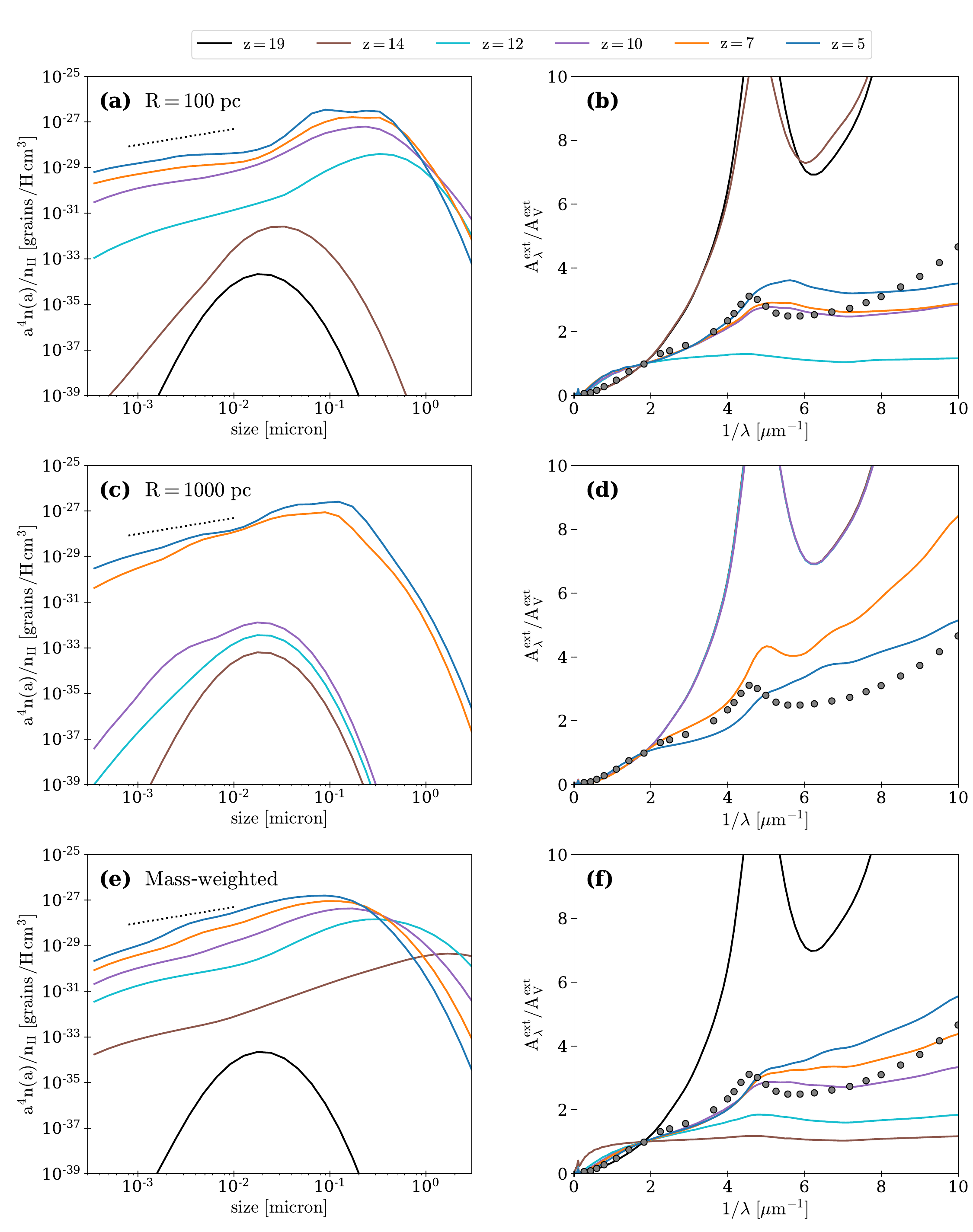}
\caption{
Left panels show the dust grain size distributions, and right panels show the corresponding extinction curves derived from our fiducial model.
The plotted size distribution is expressed as $a^4 n(a)/n_{\rm H}$, which represents the dust mass density per hydrogen nucleus.
Black, brown, cyan, purple, orange, and blue curves correspond to $z = 19, 14, 12, 10, 7,$ and $5$, respectively.
Panels (a) and (b) present the results at $R = 100~\pc$, while panels (c) and (d) show those at $R = 1000~\pc$.
Panels (e) and (f) display the grain size distributions and extinction curves averaged over the galactic disk, weighted by the dust surface mass density.
For comparison, the Milky Way grain size distribution, $n(a) \propto a^{-3.3}$ \citep{Jones1996ApJ}, is shown by the black dotted lines in panels (a), (c), and (e).
The Milky Way extinction curve with $R_{\rm V} = 3.1$ \citep[][]{Pei1992ApJ} is shown by gray circles in panels (b), (d), and (f).
}
\label{fig:hist1}
\end{figure*}

\subsection{Model parameters and fiducial setup}\label{sec:para}

Table~\ref{table:param} summarizes the key parameters in our dust evolution model.
The model predictions vary with the choice of these parameters.
In this paper, we focus on four parameters that strongly affect the results:
$\mhfive$ (the dark matter halo mass at $z=5$), $\fcm$ (the CMG mass fraction), $\yd$ (the dust yield per SN~II), and $a_0$ (the mean size of dust produced by SN~II).

When exploring the properties of individual galaxies, we adopt $\mhfive = 10^{13}~\msun$,
as such massive halos are expected to host galaxies at the bright end of the UV luminosity function.
The corresponding halo masses and number densities ($M_{\rm h},~{\rm d}n/{\rm dln}M_{\rm h}$) are approximately 
($7 \times 10^{10}~\msun,~7 \times 10^{-6}~{\rm Mpc^{-3}}$) at $z = 12$, 
($2 \times 10^{11}~\msun,~4 \times 10^{-6}~{\rm Mpc^{-3}}$) at $z = 10$, and 
($2 \times 10^{12}~\msun,~9 \times 10^{-7}~{\rm Mpc^{-3}}$) at $z = 7$ \citep{Sheth2002MNRAS}.

We adopt a fiducial CMG mass fraction of $\fcm = 0.9$, which is significantly higher than the typical values of $\fcm \sim 0.1$--0.5 inferred for local star-forming galaxies \citep[e.g.,][]{Heiles2003ApJ, Saintonge2011MNRAS}.
This choice is motivated by semi-analytic studies predicting that high-redshift galaxies can become strongly dominated by cold molecular gas.
For example, \citet{Lagos2011MNRAS} predicted $\fcm \sim 0.9$ at $z \sim 8$ \citep[see also][]{Obreschkow2009ApJ, Popping2014MNRAS}.
Such high CMG fractions naturally arise from efficient molecular gas formation driven by the high gas densities and pressures expected in high-$z$ galaxies.
Indeed, \citet{Blitz2006ApJ} derived an empirical relation between the molecular-to-atomic hydrogen mass ratio and the ISM pressure using nearby star forming galaxies:
\begin{eqnarray}
R_{\rm mol} \equiv \frac{\Sigma_{\rm H_2}}{\Sigma_{\rm HI}} \approx \left( \frac{P/k_{\rm B}}{4.3 \times 10^4~{\rm cm^{-3}~K}} \right)^{0.92} \ .
\label{eq:Blitz}
\end{eqnarray}
Our model predicts that the WNG reaches gas number densities of $\nw \gtrsim 10^2~\rm cm^{-3}$ within the half-SFR radius (see the bottom panel of Figure~\ref{fig:RP_nzt}). 
Under the assumption of $\Tw = 8000~\kelvin$, this corresponds to gas pressure of $P/k_{\rm B} \sim 8 \times 10^5~{\rm cm^{-3}~K}$.
Substituting this value into Eq.~(\ref{eq:Blitz}) yields a molecular fraction of $\fcm = (1+R_{\rm mol}^{-1})^{-1} \gtrsim 0.95$, indicating that highly molecule-dominated ISM conditions are naturally expected in the dense inner regions of our model galaxies.
On the other hand, the applicability of the empirical relation derived by \citet{Blitz2006ApJ} to galaxies at $z > 5$ remains uncertain.
We therefore explore a broad range of CMG mass fractions, $\fcm = 0.1$--$1$, throughout this paper.

The fiducial values of the dust yield per SN~II and the characteristic grain size of stellar dust are set to $(\yd,~a_0) = (10^{-4}~\msun,~10^{-2}~\mu{\rm m})$.
These values are motivated by the recent calculations of \citet{Otaki2026A&A}, who investigated dust production in SNe II under dense ISM conditions.
They showed that newly synthesized dust can be efficiently destroyed by reverse shocks before mixing with the ambient ISM, and that this destruction becomes increasingly effective at higher ambient gas densities.
In particular, they predicted dust yields of $\yd < 10^{-3}~\msun$ for ISM densities of $n_{\rm H} \gtrsim 1~{\rm cm^{-3}}$, which are several orders of magnitude smaller than the canonical yields of $\yd \sim 1~\msun$ commonly adopted in previous studies \citep[e.g.,][]{Todini2001MNRAS, Bianchi2007MNRAS}.
\citet{Otaki2026A&A} also suggested that the characteristic grain sizes of stellar dust may be significantly smaller than previously assumed.
Even grains initially produced with sizes of $a \gtrsim 0.1~\mu{\rm m}$ can undergo substantial erosion by reverse shocks, resulting in characteristic sizes of $a \lesssim 0.01~\mu{\rm m}$.
This predicted grain size is approximately one order of magnitude smaller than the commonly adopted value of $a \sim 0.1~\mu{\rm m}$ \citep[e.g.,][]{Nozawa2007ApJ, Asano2013EP&S, Hirashita2019MNRAS}.
Since the properties of stellar dust produced by SNe II remain highly uncertain, we explore a broad parameter range of $\yd = 10^{-6}$--$10^{-2}~\msun$ and two characteristic grain sizes, $a_0 = 0.01$ and $0.1~\mu{\rm m}$.

We note that the fiducial parameter set successfully reproduces various observed properties of galaxies at $z \geq 5$, as demonstrated in \S~\ref{sec:results} and \S~\ref{sec:rad}.

\section{Results}\label{sec:results}

\subsection{Metal accretion timescale}

Before presenting the main results, we briefly note that the inclusion of dense CMG clouds can significantly accelerate dust growth via metal accretion.
The top panel of Figure~\ref{fig:RP_nzt} shows the radial profiles of the metal accretion timescale for graphite grains with $a = 0.01~\mu$m, obtained for a representative galaxy with $\mhfive = 10^{13}~\msun$ under the fiducial parameter set.
The dust growth timescale in the CMG (solid curves) is approximately an order of magnitude shorter than that in the WNG (dashed curves), and is typically shorter than $1~\rm Myr$ within the half-SFR radius ($R_{50}$), nearly independent of redshift.

To clarify the physical origin of this rapid accretion, 
the bottom panel of Figure~\ref{fig:RP_nzt} shows the corresponding radial profiles of the gas number density and carbon abundance.
At $R < R_{50}$, even the WNG exhibits high gas densities of $n_{\rm H,w} \gtrsim 10^2~{\rm cm^{-3}}$. 
This naturally arises from the high gas-to-stellar mass ratios and compact sizes of galaxies at $z > 5$. 
Indeed, since the half-SFR radius generally scales as $R_{50} \propto (1+z)^{-1}$, 
the surface gas mass densities of galaxies at $z \sim 10$ become more than two orders of magnitudes higher than those of local galaxies with comparable stellar masses (see \citetalias{Toyouchi2025MNRAS}). 
Then, the CMG density in these inner region reaches $n_{\rm H,c} \gtrsim 10^4~{\rm cm^{-3}}$ under the assumption of $\fcm = 0.9$ and $\fcv = 0.01$. 
In addition, the inner disk regions exhibit moderately high carbon abundances of $Z_C \gtrsim 0.1~Z_{C,\odot}$ even at $z \sim 12$.
As a result, the densities of carbon (and silicon) in the CMG become sufficiently high to yield such short accretion timescales, as described by Eq.~(\ref{eq:tauacc}).

On the other hand, the outer disk regions ($R > 2R_{50}$) exhibit $\tau_{\rm acc} \gtrsim 10$ Myr even in the CMG phase, indicating that dust growth is highly inefficient beyond the half-SFR radius.
As discussed in more detail below, such radial variations in $\tau_{\rm acc}$ within the two-phase ISM play a crucial role in shaping dust growth across galactic disks, although their interplay with other processes, such as dust production by SNe, shattering, coagulation, and dust destruction, also contributes to the overall dust evolution.

\begin{figure*}
\centering
\includegraphics[width=1.8\columnwidth]{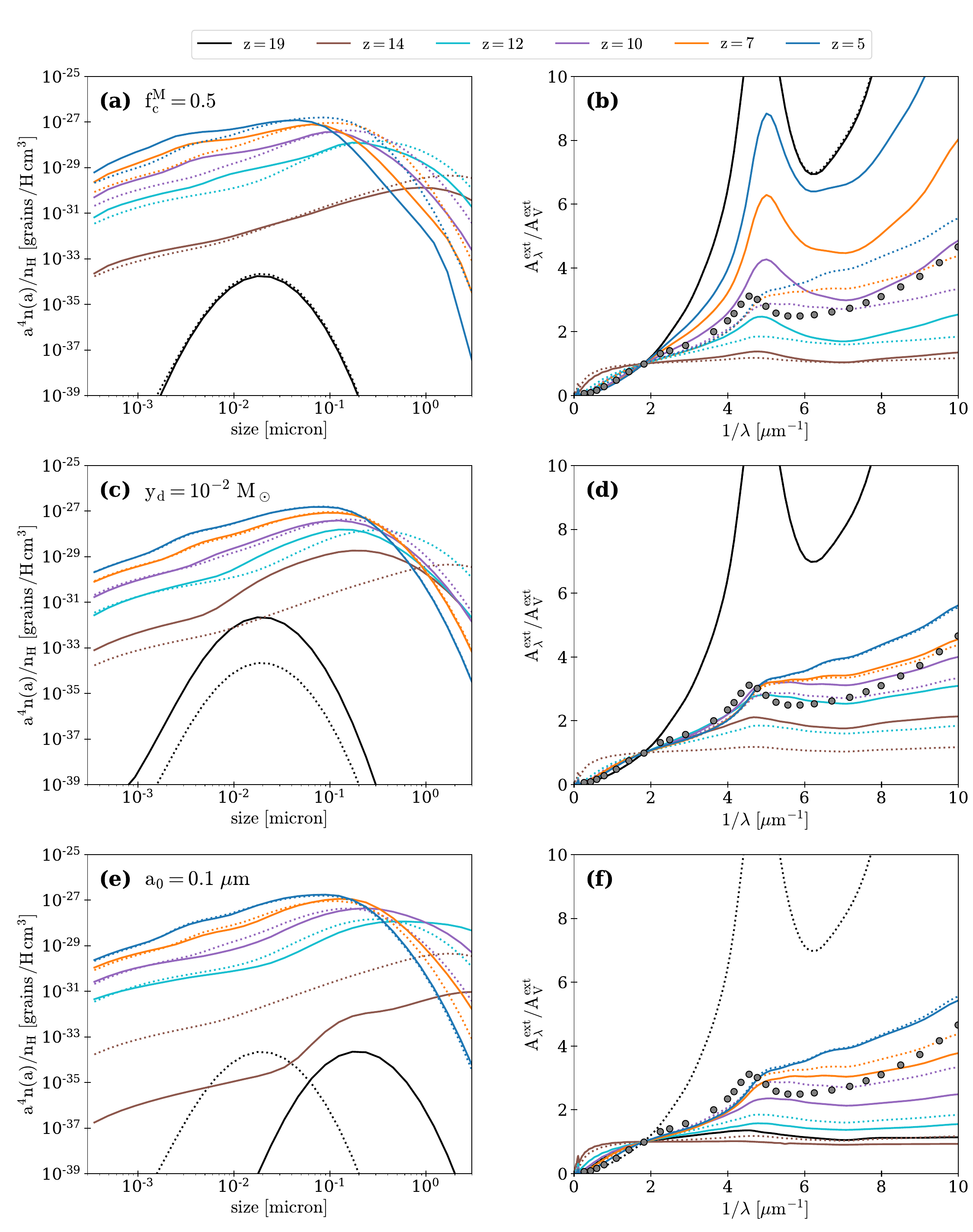}
\caption{
Same as Figures~\ref{fig:hist1}e and ~\ref{fig:hist1}f, but for models with different parameter sets from the fiducial case.
The left and right panels show the grain size distributions and extinction curves averaged over the galactic disk, weighted by the dust surface mass density.
Panels (a) and (b) present the results for the model with $\fcm = 0.5$, panels (c) and (d) for $\yd = 10^{-2}~\msun$, and panels (e) and (f) for $a_0 = 0.1~\mu$m.
For comparison, the results of the fiducial model are shown by dotted lines in all panels.
}
\label{fig:hist2}
\end{figure*}

\subsection{Size distribution and extinction curve}\label{sec:size}

Figure~\ref{fig:hist1} shows the grain size distributions (left column) and extinction curves (right column) obtained in the fiducial model,
where different colored curves correspond to results at redshifts $z = 5$--19.
For the size distributions, we plot $a^4 n(a)/n_{\rm H}$ to highlight which grain sizes contribute most to the total dust mass.

Panels (a) and (b) present the results at a small radius of $R = 100~\pc$, which lies inside the half-SFR radius at $z \lesssim 18$.
At $z = 19$, dust supplied by SNe II dominates the dust budget, and the size distribution follows a lognormal distribution with a peak at $a \sim 0.01~\mu{\rm m}$ (see Eq.~\ref{eq:lognormal}).
As evolution proceeds, the size distribution develops a tail toward smaller sizes due to shattering in the WNG region, which produces small grains with $a \sim 10^{-3}~\mu{\rm m}$.
The short metal accretion timescale in the CMG phase, $\tau_{\rm acc} \lesssim 1~\Myr$ for these small grains (see Figure~\ref{fig:RP_nzt}), drives rapid dust growth and increases the total dust mass density in the ISM.
Since metal accretion also increases grain sizes, the size distribution becomes dominated by large grains with $a \gtrsim 0.1~\mu{\rm m}$.
From $z \sim 10$ to 5, however, the fraction of very large grains ($a \gtrsim 0.5~\mu{\rm m}$) gradually decreases with time.
This is because C and Si in the ISM are significantly depleted by metal accretion, reducing the efficiency of further dust growth,
so that dust destruction by supernovae becomes dominant for these large grains.

The extinction curve evolves in response to this change in the grain size distribution.
At $z = 19$, the curve is extremely steep and exhibits a prominent bump at $\lambda \sim 2000$~\AA, reflecting the dominance of $a \sim 0.01~\mu{\rm m}$ grains.
At $z \sim 12$, large grains with $a \gtrsim 0.1~\mu{\rm m}$ absorb radiation almost uniformly at wavelengths $\lambda \lesssim 2 \pi a$, making the extinction curve relatively flat from the optical to the UV.
At later times, the destruction of large grains by supernovae increases the relative abundance of small grains, gradually steepening the extinction curve again.
By $z = 5$, the extinction curve becomes comparable to the Milky Way extinction curve shown by the gray points.

Panels (c) and (d) show the results at a larger radius of $R = 1000~\pc$, which remains outside the half-SFR radius until $z \sim 8$.
In this region, shattering first produces small grains, and large grains gradually become dominant through metal accretion and coagulation, similar to the behavior at $R = 100~\pc$.
However, because the gas density is significantly lower than in the inner region, dust growth proceeds more slowly.
The fraction of large grains with $a > 0.1~\mu{\rm m}$ and the total dust abundance rapidly increases between $z = 10$ and 7,
because the half-SFR radius expands beyond $R = 1000~\pc$, leading to a significant decease of the metal accretion timescale, as shown in Figure~\ref{fig:RP_nzt}.
Nevertheless, the fraction of small grains with $a < 0.1~\mu{\rm m}$ remains higher than in the inner region, and the extinction curve is correspondingly steeper, reflecting the overall slower dust evolution.

To characterize the global dust properties of the galaxy, we average the grain size distributions and extinction curves over the entire galactic disk, weighted by the dust surface mass densities ($\Sigma_{\rm d} \equiv D_{\rm tot}\sgmg$).
Panels (e) and (f) show the averaged size distributions and extinction curves.
Reflecting the inside-out growth of the galaxy, these averaged quantities initially resemble those of the inner regions, and gradually evolve toward those of the outer regions.
At $z = 5$, the grain size distribution peaks around $a \sim 0.1~\mu$m and exhibits a power law tail toward smaller size with $n(a) \propto a^{-3}$, which is slightly shallower than that inferred for the Milky Way, $n(a) \propto a^{-3.3}$ \citep{Jones1996ApJ}.
The resulting extinction curve is roughly consistent with, or slightly steeper than, that of the Milky Way.

Next, we investigate the dependence of the results on the model parameters governing dust size evolution.
Figure~\ref{fig:hist2} shows the time evolution of the grain size distributions and the extinction curves for three models, in which the parameters $\fcm$, $\yd$, and $a_0$ are varied from the fiducial values.
The quantities shown here are averages weighted by $\Sigma_{\rm d}$, in the same manner as in panels (e) and (f) of Figure~\ref{fig:hist1}.

Panels (a) and (b) of Figure~\ref{fig:hist2} show the results for a model adopting a CMG mass fraction of $\fcm = 0.5$, which is about a factor of two smaller than in the fiducial model.
Compared with the fiducial model, shown by the dotted lines, this model exhibits a significantly smaller fraction of large grains with $a > 0.1~\mu{\rm m}$.
This is because the amount of CMG, which is the primary site of metal accretion and coagulation responsible for producing large grains, is reduced, while the fraction of WNG, where shattering produces small grains, increases.
Reflecting this small-grain-dominated size distribution, the resulting extinction curve is much steeper than that of the fiducial model.

Panels (c) and (d) show a model assuming a dust yield of $\yd = 10^{-2}~\msun$, which is two orders of magnitude larger than in the fiducial model.
In this case, the larger dust abundance in the ISM leads to a faster evolution of the grain size distribution compared to the fiducial model.
This difference is particularly pronounced at $z \gtrsim 12$.
At $z < 10$, however, the dust properties become dominated by interstellar processes such as metal accretion and shattering.
As a result, the grain size distribution and the resulting extinction curve become nearly identical to those of the fiducial model.

Finally, Panels (e) and (f) show a model adopting a larger SN dust size of $a_0 = 0.1~\mu{\rm m}$, which is ten times larger than in the fiducial model.
In this case, the dust mass density distribution at $z = 14$ is substantially lower than in the fiducial model over the entire grain-size range,
indicating significantly less efficient dust growth in the early phase.
This is because small grains with $a \sim 10^{-3}~\mu{\rm m}$, which act as the primary seeds for metal accretion, are scarcely produced by supernovae in this model, delaying the onset of dust growth.
Consequently, the grain size distribution remains biased toward larger grains compared with the fiducial model until $z \sim 7$.
Consistent with this trend, the extinction curve is flatter than that of the fiducial model.
However, as in the case of $\yd = 10^{-2}~\msun$, the difference from the fiducial model gradually decreases with time, and the results become broadly consistent by $z \sim 5$.

\begin{figure*}
\centering
\includegraphics[width=2.0\columnwidth]{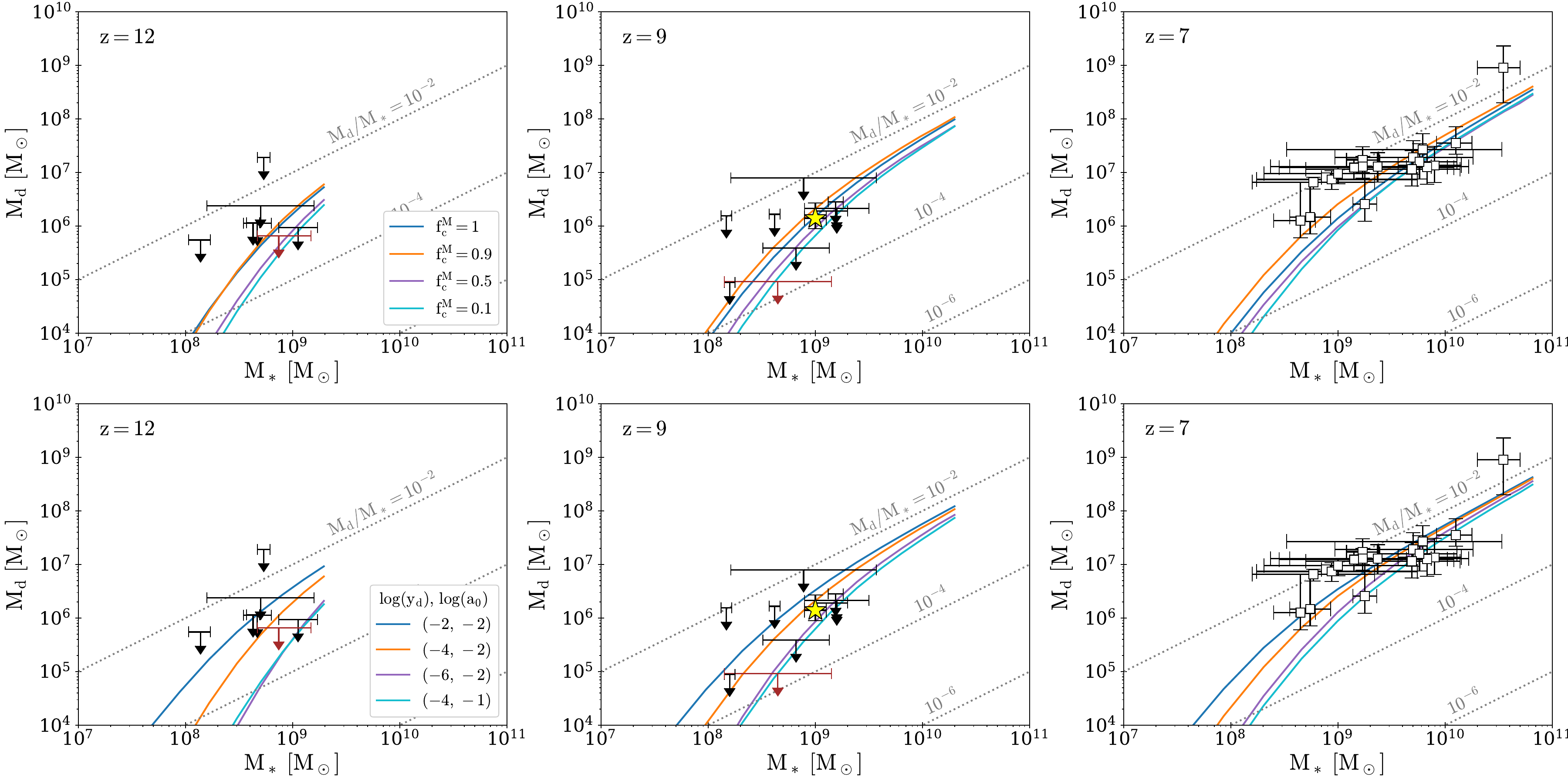}
\caption{
Relation between dust and stellar masses obtained from our calculations for 21 dark matter halos.
The left, middle, and right panels show the results at $z = 12$, 9, and 7, respectively.
In the upper row, models with different CMG mass fractions are compared: $\fcm = 1$ (blue), $0.9$ (orange), $0.5$ (purple), and $0.1$ (cyan).
In the lower row, models with different SN dust properties are compared: $(\log(\yd/\msun),\,\log(a_0/\mu{\rm m})) = (-2,\,-2)$ (blue), $(-4,\,-2)$ (orange), $(-6,\,-2)$ (purple), and $(-4,\,-1)$ (cyan).
In both the upper and lower panels, the orange lines represent the fiducial case in Table~\ref{table:param}.
For comparison, observational data for galaxies at the corresponding redshift ranges are also shown.
White squares in the right panels represent galaxies at $6 < z < 8$ taken from \citet{Hashimoto2019PASJ, Reuter2020ApJ, Bakx2021MNRAS, Sommovigo2022MNRAS, Umehata2025arXiv}.
The yellow star in the middle panels represents the observational data of MACS0416\_Y1 at $z = 8.31$, currently the highest redshift galaxy with a dust mass measurement \citep{Bakx2025MNRAS}.
In the left and middle panels, black downward arrows indicate upper limits on the dust masses of individual galaxies at $8 < z < 10$ and $z > 10$, inferred by \citet{Algera2023MNRAS, Bakx2026MNRAS}.
Brown downward arrows correspond to upper limits on the dust masses inferred from the stacking analysis for galaxies at $z < 9.5$ and $z > 9.5$ \citep{Bakx2026MNRAS}.
Here, we include only spectroscopically confirmed galaxies from the literature.
Accordingly, we exclude two photometrically identified galaxies, SPT-0615 ($z_{\rm ph} = 9.63$) and COS-z12-1 ($z_{\rm ph} = 12.25$), which were reported to have extremely low dust-to-stellar mass ratios of $M_{\rm d}/M_\ast < 3.2 \times 10^{-4}$ and $< 4.3 \times 10^{-4}$, respectively \citep{Bakx2026MNRAS}.
}
\label{fig:Md_z}
\end{figure*}

\subsection{Dust mass}\label{sec:mass}

Metal accretion onto dust grains in the ISM is the key process that increases both grain sizes and masses.
Therefore, the total dust mass of galaxies is expected to increase in tandem with the evolution of grain size distribution.

Figure~\ref{fig:Md_z} shows the relation between dust mass ($M_{\rm d}$) and stellar mass ($M_{\ast}$), obtained from our calculations for 21 dark matter halos spanning $\mhfive = 10^{10}$--$10^{14}~\msun$.
The left, middle, and right panels correspond to the results at $z = 12$, 9, and 7, respectively.
For comparison, we also show observational data for galaxies identified at the corresponding redshift ranges \citep[][]{Hashimoto2019PASJ, Reuter2020ApJ, Sommovigo2022MNRAS, Bakx2021MNRAS, Bakx2025MNRAS, Bakx2026MNRAS}.
We further highlight the dependence on $\fcm$ and on the properties of SN dust ($\yd$, $a_0$)
in the upper and lower rows, respectively.

We first focus on the fiducial model, shown by the orange curves in all panels.
From the left to right panels, the overall shape of the predicted $M_{\rm d}$-$M_\ast$ relation evolves only weakly with redshift.
In contrast, the dust-to-stellar mass ratio increases systematically with stellar mass.
Low-mass galaxies with $M_\ast \lesssim 10^8~\msun$ have $M_{\rm d}/M_\ast < 10^{-4}$,
whereas massive galaxies with $M_\ast \gtrsim 10^9~\msun$ show dust-to-stellar mass ratios converging to $M_{\rm d}/M_\ast \sim 10^{-2}$.
This behavior suggests that rapid dust growth occurs once galaxies reach $M_\ast \sim 10^8$--$10^9~\msun$.
Such a critical stellar mass naturally explains the prevalence of dust-poor galaxies observed at $z > 10$, since the most of these galaxies at these epochs remain less massive than $M_\ast \sim 10^9~\msun$ \citep[but also][]{Harikane2026arXiv}.

Our fiducial model is broadly consistent with the observations of massive galaxies at $z \sim 7$.
Furthermore, it successfully reproduces both the dust and stellar masses of MACS0416\_Y1 at $z = 8.31$, currently the highest redshift galaxy with a detected dust thermal emission \citep{Bakx2025MNRAS}, indicated by the yellow star in the middle panels.
This result is also consistent with the findings of \citet{Kano2026arXiv}, who calculated full grain-size evolution using a one-zone galaxy model.
They demonstrated that a high ISM density of $n_{\rm H} \sim 10^4~{\rm cm^{-3}}$ is required to reproduce the SED of MACS0416\_Y1.
A key result of our calculations is that such high gas densities naturally emerge once both the size evolution of galaxies and the presence of a two-phase ISM are taken into account.

It is also worth noting that our theoretical predictions do not significantly violate the upper limits on dust masses derived for individual galaxies at $z > 9$ (black arrows) and those inferred from stacking analyses (red arrows) by \citet{Bakx2026MNRAS}.
Although the fiducial model slightly overestimates the dust mass relative to the upper limit for GHZ2, which has $M_{\rm d}/M_\ast < 8.3 \times 10^{-5}$ at $z = 12.33$, this galaxy may contain a non-negligible AGN contribution to its UV emission, potentially leading to an overestimation of the stellar mass \citep{Calabro2024ApJ}.
The more conservative estimate of $\log_{10}M_\ast = 8.3$ reported by \citet{Zavala2025NatAs} instead gives $M_{\rm d}/M_\ast < 4.7 \times 10^{-4}$,
with which our fiducial model remains consistent.


Next, we discuss the dependence on the CMG mass fraction, shown in the upper panels of Figure~\ref{fig:Md_z}.
We find that the fiducial model with $\fcm = 0.9$ yields the largest dust mass, and that the dust mass generally decreases with decreasing $\fcm$.
This indicates that a higher abundance of CMG, where metal accretion occurs, promotes efficient dust growth.
The slight reduction in dust mass for $\fcm = 1$ compared to the fiducial model can be attributed to the absence of WNG, where shattering produces small grains that act as seeds for metal accretion.
We also note that the model with $\fcm = 0.1$ yields a comparable dust mass to the $\fcm = 0.5$ model at $z \sim 5$.
This is because, in the $\fcm = 0.1$ case, the WNG still maintains a relatively high density, allowing both shattering and metal accretion to proceed efficiently even without a significant CMG component.

Finally, we discuss the dependence on the SN dust properties, shown in the lower panels of Figure~\ref{fig:Md_z}.
We find a strong dependence on $\yd$ and $a_0$, particularly at higher redshifts.
The model with $\yd = 10^{-2}~\msun$ exceeds the observational upper limits of the dust mass for several individual galaxies at $z > 8$, as well as the upper limit from the stacking analysis.
This implies an upper bound on the SN dust yield of $\yd \lesssim 10^{-3}~\msun$.
On the other hand, even an extremely small dust yield of $\yd = 10^{-6}~\msun$ can be consistent with the observations at $z \sim 7$, indicating rapid dust growth within the ISM.
For a larger SN dust size of $a_0 = 0.1~\mu{\rm m}$, the increase in dust mass is significantly delayed.
This is due to inefficient metal accretion caused by the lack of small grains, as shown in Figure~\ref{fig:hist2}e.

This result about the SN dust properties provides an important implication for dust growth in high-redshift galaxies.
Even if, as suggested by \citet{Otaki2026A&A}, strong reverse shocks in dense ISM environments reduce the characteristic size of SN-produced dust to values about one order of magnitude smaller than previously assumed \citep[e.g.,][]{Nozawa2007ApJ, Hirashita2019MNRAS}, such small stellar grains are not necessarily disadvantageous for subsequent dust evolution.
Instead, they can efficiently act as seeds for metal accretion and thereby accelerate rapid dust growth in the ISM.
Thus, the negative effect of lower dust yields caused by strong reverse shocks can, at least partly, be compensated by the accompanying reduction in grain sizes.

Overall, these results demonstrate that the dust mass growth in high-redshift galaxies is influenced by both the physical conditions of the ISM and the properties of SN-produced dust.

\begin{figure}
\centering
\includegraphics[width=0.9\columnwidth]{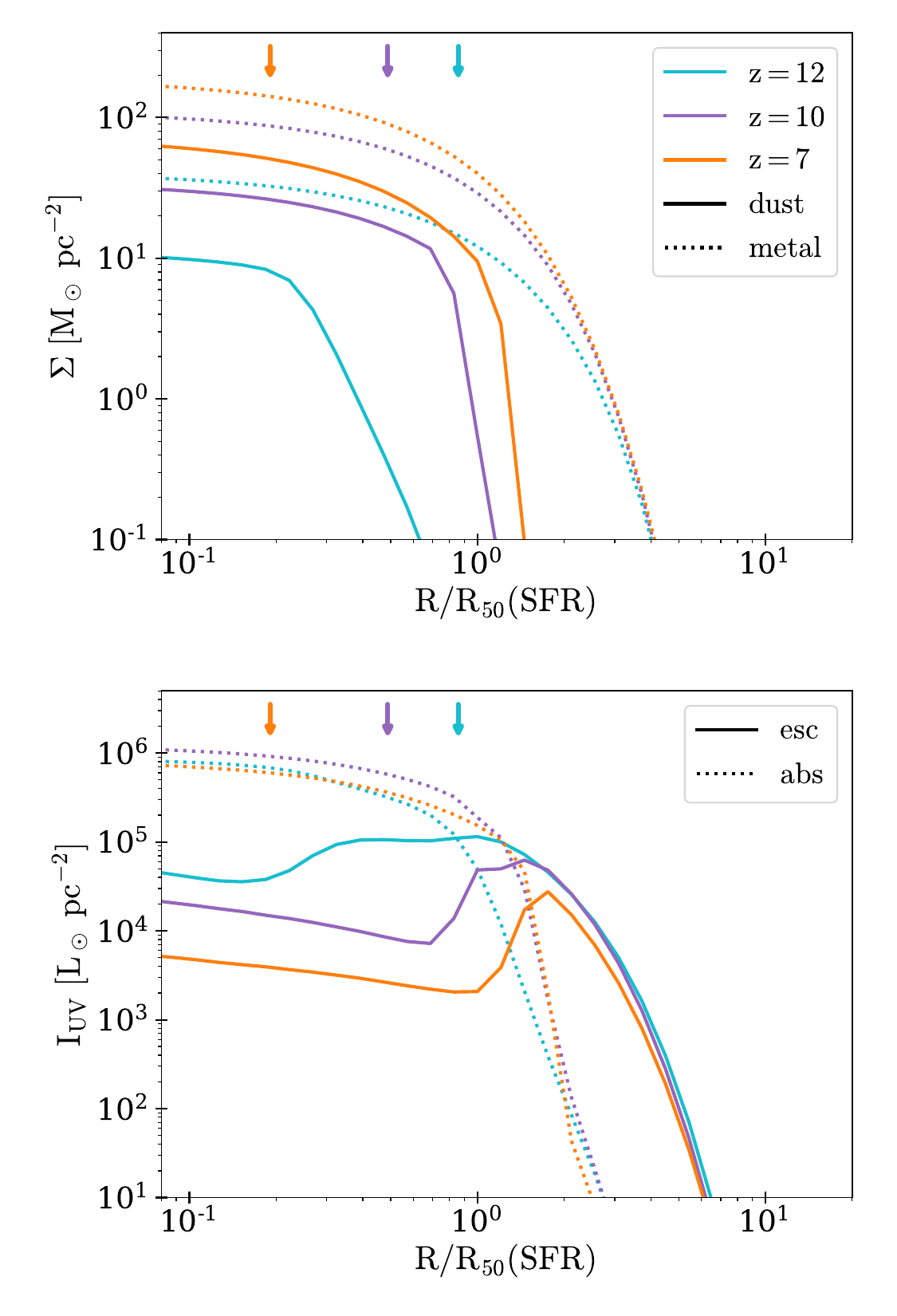}
\caption{
Top panel shows the radial profiles of the surface mass densities of dust (solid lines) and metals (dotted lines) for the galaxy with $\mhfive = 10^{13}~\msun$ obtained in the fiducial model.
Bottom panel shows the radial profiles of the UV surface brightness for the same galaxy, where solid and dotted lines represent the escaping and absorbed UV radiation, respectively.
In the both panel, the horizontal axis is normalized by half-SFR radius.
Cyan, purple, orange, and blue curves correspond to $z = 12$, $10$, $7$, and $5$, respectively.
Downward arrows indicate a physical scale of 200~pc, which roughly corresponds to the spatial resolution of JWST at the corresponding redshifts.
}
\label{fig:RP}
\end{figure}

\begin{figure}
\centering
\includegraphics[width=0.9\columnwidth]{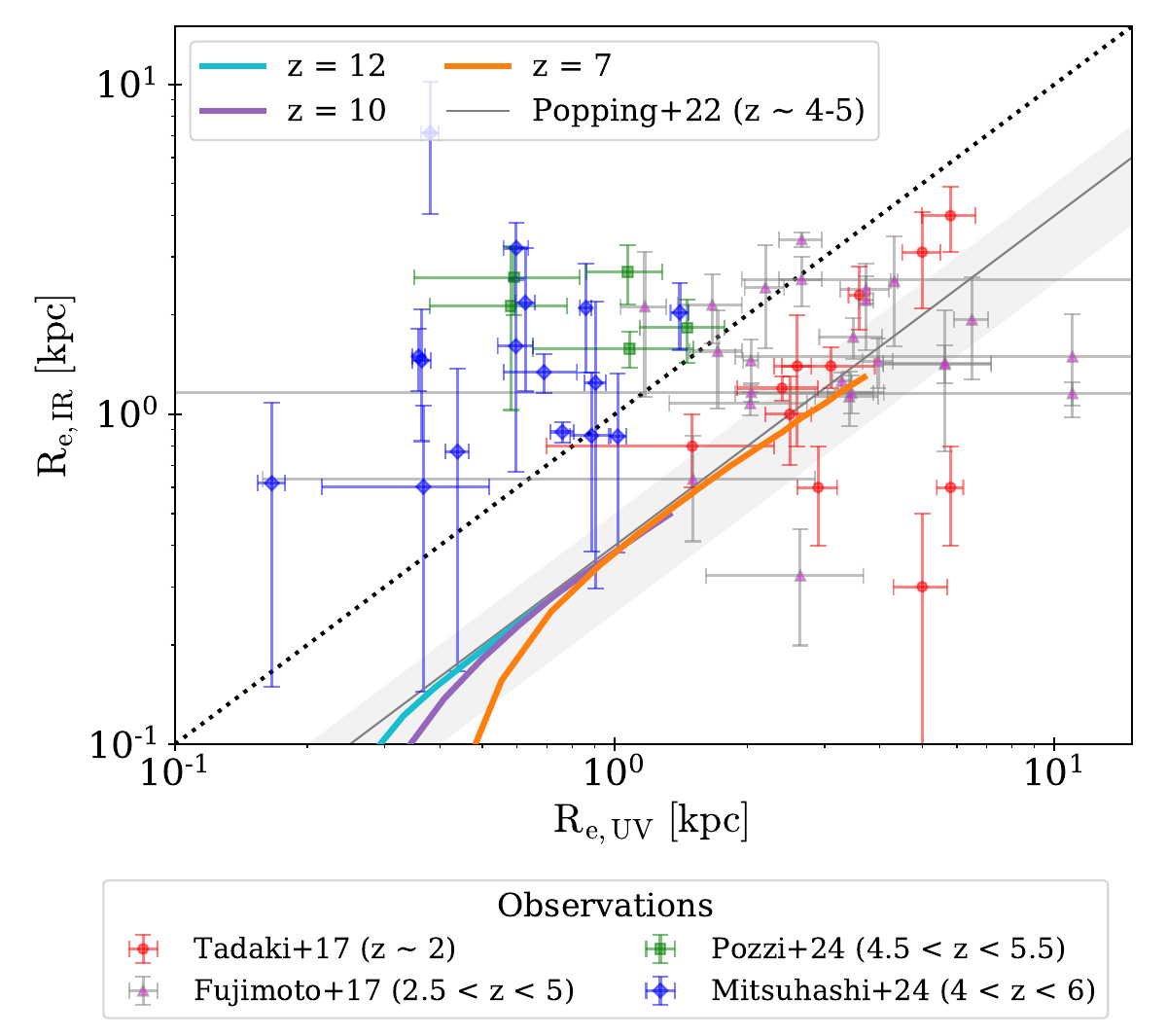}
\caption{
Relation between the UV and IR effective radii, $R_{\rm e,UV}$ and $R_{\rm e,IR}$, respectively, obtained from our calculations for 21 dark matter halos under the fiducial parameter set.
The cyan, purple, and orange curves represent the results at $z = 12$, 10, and 7, respectively.
The black dotted line corresponds to $R_{\rm e,UV} = R_{\rm e,IR}$.
For comparison, the gray solid line and shaded region indicate the theoretical prediction for galaxies at $z \sim 4$-5 from the Illustris-TNG50 simulations and its $1\sigma$ dispersion, respectively \citep{Popping2022MNRAS}.
Different symbols show observational data taken from 
\citet{Tadaki2017ApJ} (red circles), 
\citet{Fujimoto2017ApJ} (magenta triangles), 
\citet{Pozzi2024A&A} (green squares), 
and \citet{Mitsuhashi2024A&A} (blue diamonds).
}
\label{fig:Reff}
\end{figure}

\subsection{Radial distribution of dust}\label{sec:radial}

The top panel of Figure~\ref{fig:RP} shows the radial profiles of the dust and metal surface mass densities, $\Sigma_{\rm d}$ and $\Sigma_{\rm Z}$, respectively, for a galaxy with $\mhfive = 10^{13}~\msun$ in the fiducial model.
Here, the horizontal axis is normalized by the half-SFR radius ($R_{50}$).
Reflecting the inside-out growth of the galaxy, $\Sigma_{\rm d}$ gradually extends to larger radii with time,
but remains significantly more centrally concentrated than $\Sigma_{\rm Z}$ at all redshifts.
This is because, in the outer regions of the galaxy, the gas density is still too low for efficient dust growth via metal accretion,
as shown in Figure~\ref{fig:RP_nzt}.

This radial distribution of dust strongly affects the UV radiation emerging from the galactic disk.
The bottom panel of Figure~\ref{fig:RP} shows the radial profiles of the UV surface brightness,
where the solid and dashed lines represent the escaping and absorbed UV radiation, respectively.
In the inner regions ($R < R_{50}$), where dust is abundant,
most of the UV radiation is absorbed, yielding $I^{\rm esc}_{\rm UV}/I^{\rm abs}_{\rm UV} \sim 10^{-2}$.
In contrast, the outer regions ($R > R_{50}$) remain highly transparent to UV radiation, reflecting the delayed dust growth relative to star formation.
Consequently, even if the inner regions become extremely faint in the UV due to dust extinction, a non-negligible fraction of the total UV luminosity still originates from the outer regions, where dust extinction is negligible.
This effect contributes to shaping the extended bright-end of the UV luminosity function at $z \sim 7$, where galaxies are already dust-rich (see \S~\ref{sec:rad}).
These results highlight that resolving the radial density structure of galaxies is crucial for understanding their integrated UV luminosities and dust extinction properties.

Our model demonstrates that high-$z$ galaxies exhibit UV-dark holes in their central regions.
To discuss the observational feasibility of these predicted dark holes, the downward arrows in Figure~\ref{fig:RP} indicate a physical scale of 200~pc, which roughly corresponds to the spatial resolution of JWST at the corresponding redshifts.
At $z \gtrsim 12$, the dark holes are not sufficiently extended to be spatially resolved by JWST,
whereas they become large enough to be identified at $z \lesssim 10$.
Indeed, such dark-hole-like structure have been reported in some dusty galaxies at $z \sim 7$ \citep{Schouws2022ApJ, Rowland2024MNRAS},
as well as in star-forming galaxies at lower redshifts \citep[e.g,][]{Rujopakarn2016ApJ, Elbaz2018A&A, Gomez2022A&A}.
These observational results support our model predictions.
Nevertheless, the ubiquitous existence of dusty galaxies without clear UV-dark holes at $z \lesssim 7$ may indicate that our model require additional mechanisms to effectively reduce optical depth in the central regions of galaxies, as discussed in \S~\ref{sec:caveats}.

In addition, our model predicts that dust thermal emission in the IR is more compact than UV emission in galaxies.
Figure~\ref{fig:Reff} shows the relation between the UV and IR effective radii of galaxies, $R_{\rm e,UV}$ and $R_{\rm e,IR}$, respectively, at $z =$ 7, 10, and 12, calculated using the fiducial model.
The theoretical curves represent the global relation obtained from 21 sample galaxies.
We find that the IR effective radius roughly scales with the UV one and is systematically smaller by a factor of a few, i.e., $R_{\rm e,IR}/R_{\rm e,UV} \lesssim 0.5$, nearly independent of redshift.
This result is consistent with the predictions from the Illustris-TNG50 simulations by \citet{Popping2022MNRAS}, who argued that dust obscuration in the central regions of galaxies can increase the observed UV effective radii relative to the half-SFR radii, particularly at higher redshifts.
Furthermore, our model naturally explains the compact dust thermal emission observed in massive star-forming galaxies at $z = 0$-$6$ with stellar masses of $\log(M_\ast/\msun) \gtrsim 10.5$ and UV effective radii of $R_{\rm e} \gtrsim 2~\kpc$ \citep{Tadaki2017ApJ, Fujimoto2017ApJ}.

It is worth noting, however, that our model prediction is in tension with observations indicating that galaxies with $\log(M_\ast/\msun) \lesssim 10.5$ at $z =$ 4-6 are generally compact in the UV, $R_{\rm e,UV} < 2~{\rm kpc}$, but relatively extended in the IR, i.e., $R_{\rm e,IR}/R_{\rm e,UV} > 1$ \citep{Pozzi2024A&A, Mitsuhashi2024A&A}.
This observational trend implies that dusty outflows are more effective in such less massive galaxies because of their shallower gravitational potentials,
thereby extending the spatial distribution of dust.
We will further discuss the potential impact of such dust redistribution on our results in \S~\ref{sec:caveats}.

\begin{figure*}
\centering
\includegraphics[width=1.8\columnwidth]{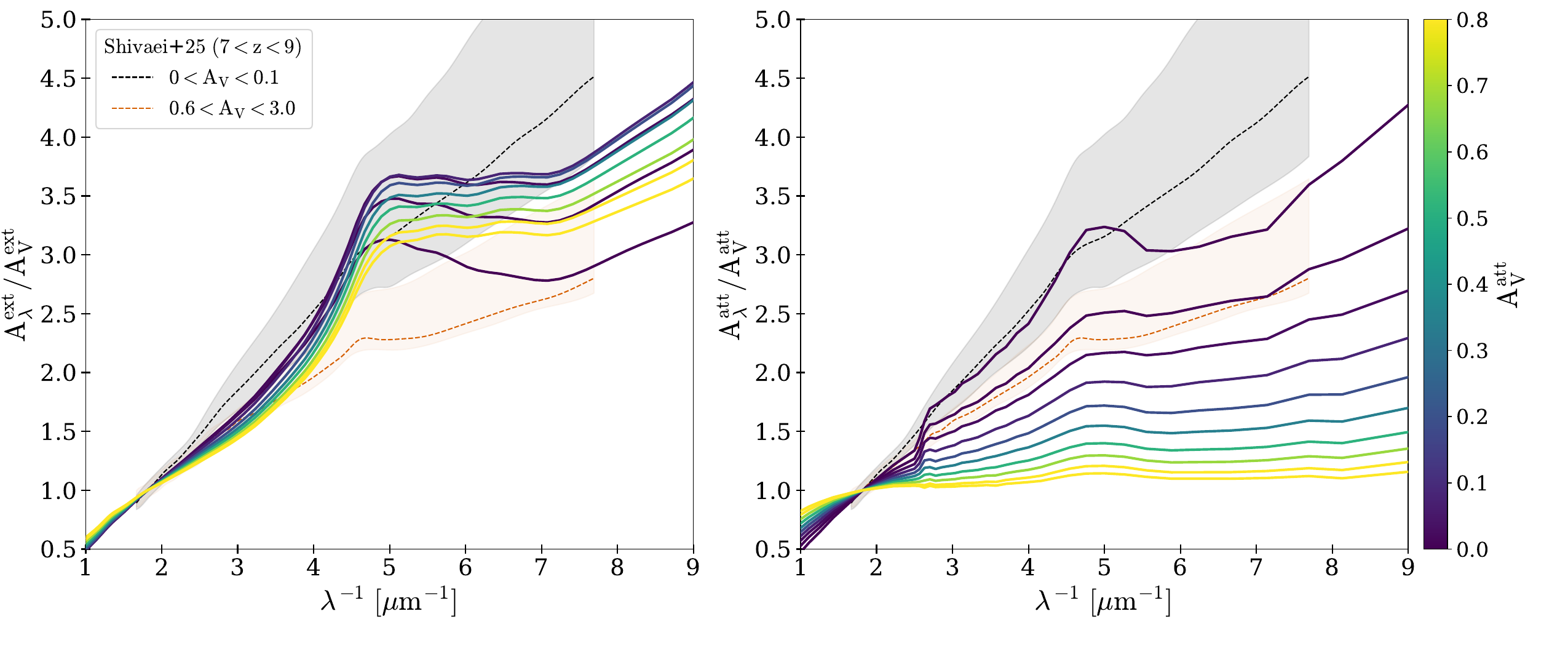}
\caption{
Extinction and attenuation curves of our model galaxies in a stellar mass range of $\mstar = 10^{7}$--$10^{10}~\msun$ at $z = 8$ are shown in the left and right panels, respectively.
The color of each curve indicates the V-band attenuation, $A^{\rm att}_{\rm V}$.
For comparison, the orange and black dotted lines represent the median attenuation curves for observed galaxies with $0 < A^{\rm att}_{\rm V} < 0.1$ and $0.6 < A^{\rm att}_{\rm V} < 3.0$, respectively, at $7 < z < 9$, reported by \citet{Shivaei2025arXiv}.
The corresponding $1\sigma$ dispersions are shown by the shaded regions.
}
\label{fig:att_curve}
\end{figure*}

\section{Radiative properties}\label{sec:rad}

In this section, we further investigate the radiative properties of our model galaxies and discuss how it is affected by the dust evolution described in \S~\ref{sec:results}.

\subsection{Attenuation curve}\label{sec:att_curve}

Attenuation curves of galaxies can be observationally inferred through spectral energy distribution fitting.
Recently, \citet{Shivaei2025arXiv} statistically derived attenuation curves, using a spectroscopic-redshift sample of $\sim$ 3,800 galaxies at $z \sim 1$--9, including 124 galaxies at $z = 7$--9.
They found that the attenuation curves systematically vary with redshift, stellar mass, and star formation rate.
Such variations are likely driven by the evolution of the grain size distribution, the total dust mass, and the relative spatial distribution between stars and dust during galaxy evolution \citep[see also][]{Markov2025NatAs, Markov2025A&A}.

Here, we first investigate the variations of extinction curves among our model galaxies, which are directly linked to the underlying grain-size distributions and therefore provide a useful basis for understanding the attenuation curves discussed later.
The left panel of Figure~\ref{fig:att_curve} shows the extinction curves of our model galaxies at $z = 8$ obtained with the fiducial model, where the color of each curve indicates the corresponding $A^{\rm att}_{\rm V}$ magnitude.
We find that the extinction curves tend to become systematically flatter with increasing $A^{\rm att}_{\rm V}$.
This trend arises because more dust-obscured galaxies possess denser and more metal-enriched ISM, where dust growth efficiently produces large-grain-dominated size distributions over wide regions of galactic disks.
An exception is found for galaxies with negligible attenuation ($A_{\rm V} \sim 0$), in which efficient dust growth occurs only in the innermost regions, and the grain-size distribution of those regions dominates the overall extinction curve of the galaxy.

The right panel of Figure~\ref{fig:att_curve} shows the corresponding attenuation curves.
We again find a flattening trend with increasing $A^{\rm att}_{\rm V}$.
Galaxies with weak attenuation ($A^{\rm att}_{\rm V} < 0.1$) exhibit steep attenuation curves with a prominent $2000$~\AA\ bump, reaching $A^{\rm att}_{1500}/A^{\rm att}_V \gtrsim 3$.
In contrast, more dust-obscured galaxies with $A^{\rm att}_{\rm V} > 0.5$ show substantially flatter attenuation curves with $A^{\rm att}_{1500}/A^{\rm att}_V \lesssim 1.5$. 
Remarkably, the flattening trend is significantly stronger than that found in the extinction curves.
This implies that the flattening trend is not driven solely by variations in the grain-size distribution, but that the spatial geometry of dust and stars within galaxies also plays a crucial role.



Here, we highlight two key factors that produce flat attenuation curves in our model.
First, our adoption of a slab-like geometry for stars and dust (Eqs.~\ref{eq:fesc_vy} and \ref{eq:fesc_old}) naturally weakens the wavelength dependence of attenuation compared to extinction.
For example, in the optically thick limit ($\tau_\lambda \gg 1$), attenuation scales only logarithmically with optical depth, $A^{\rm att}_{\lambda} \propto \log(\tau_\lambda)$, whereas extinction scales linearly as $A^{\rm ext}_{\lambda} \propto \tau_\lambda$.
As a result, attenuation curves become significantly flatter than the underlying extinction curves, particularly in highly obscured systems.

Second, the attenuation curves preferentially reflect the dust properties of the inner regions within the half-SFR radius, where dust growth proceeds most efficiently owing to high gas densities and metallicities.
These regions rapidly develop large-grain-dominated size distributions, resulting in flatter extinction curves (see Figure~\ref{fig:hist1}).
Meanwhile, the outer regions, where small grains are relatively more abundant, contribute little to the total attenuation because the dust abundance remain low.
Thus, the flat attenuation curves predicted by our model also reflect the spatial concentration of evolved dust relative to stars that is shown in \S~\ref{sec:radial}.

Finally, we compare our predictions with the observationally inferred attenuation curves of \citet{Shivaei2025arXiv}, which are shown by the gray and orange dotted curves.
While the observations also exhibit the flattening of attenuation curves with increasing $A^{\rm att}_{\rm V}$,
the observed attenuation curves are systematically steeper than our predictions.
Even relatively obscured galaxies with $A_{\rm V} > 0.5$ retain moderately steep attenuation curves with $A^{\rm att}_{1500}/A^{\rm att}_V \gtrsim 2$.
Similar trends have also been reported in other recent studies of high-redshift galaxies \citep{Fisher2025MNRAS, Markov2025NatAs, Markov2025A&A}.
This discrepancy may arise from several factors, such as the assumed dust geometry, the simplified radiative transfer treatment, and the absence of physical processes that redistribute dust within galaxies.
A more comprehensive investigation of these effects is left for future work.

\begin{figure}
\centering
\includegraphics[width=0.9\columnwidth]{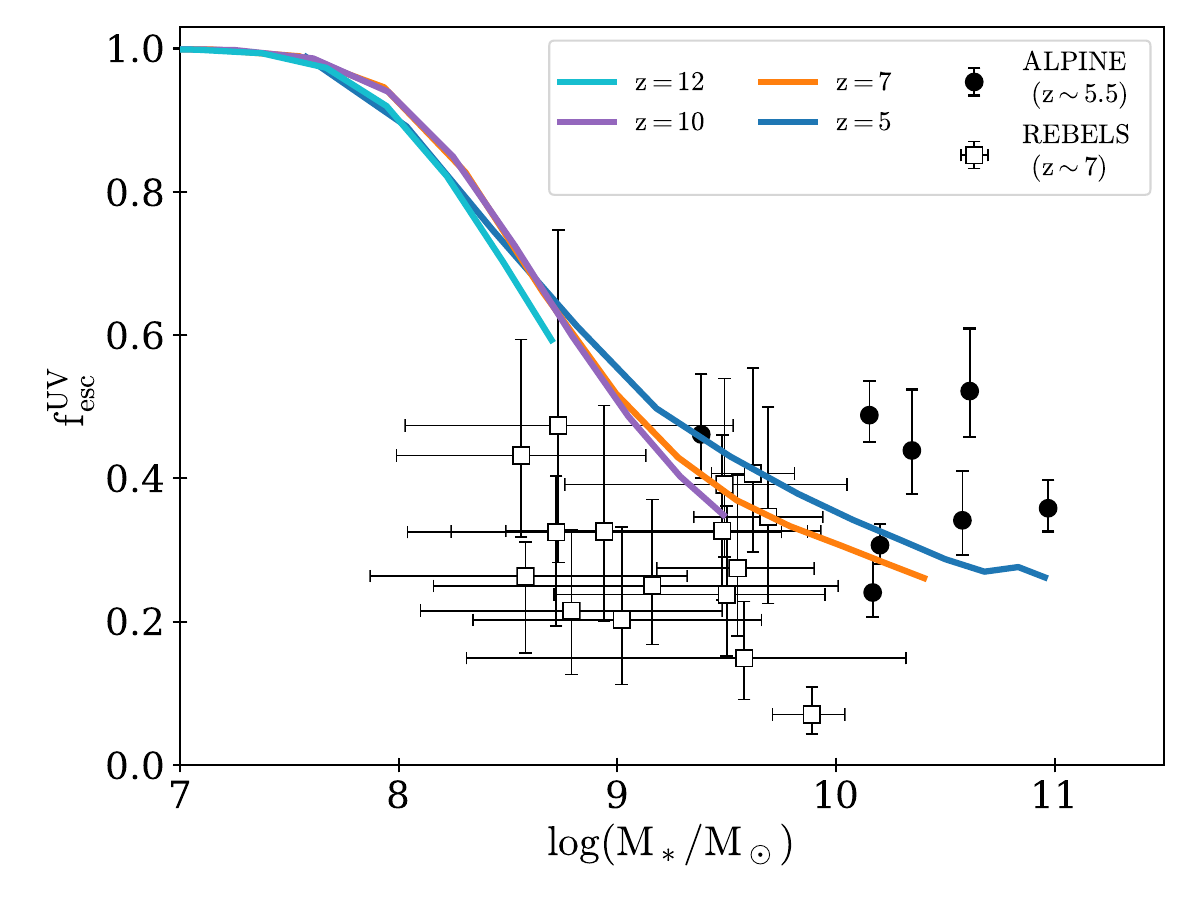}
\caption{
UV escape fraction from galaxies as a function of stellar masses obtained by our fiducial model. The cyan, purple, orange, and blue curves represent the results at $z = 12$, 10, 7, and 5, respectively.
For comparison, black circles show observational data for galaxies at $z \sim 5.5$ identified by the ALPINE survey \citep{Fudamoto2020A&A},
whereas white squares correspond to galaxies at $z \sim 7$ identified by the REBELS survey \citep{Bouwens2022ApJ, Inami2022MNRAS}.
}
\label{fig:fesc}
\end{figure}

\begin{figure}
\centering
\includegraphics[width=0.9\columnwidth]{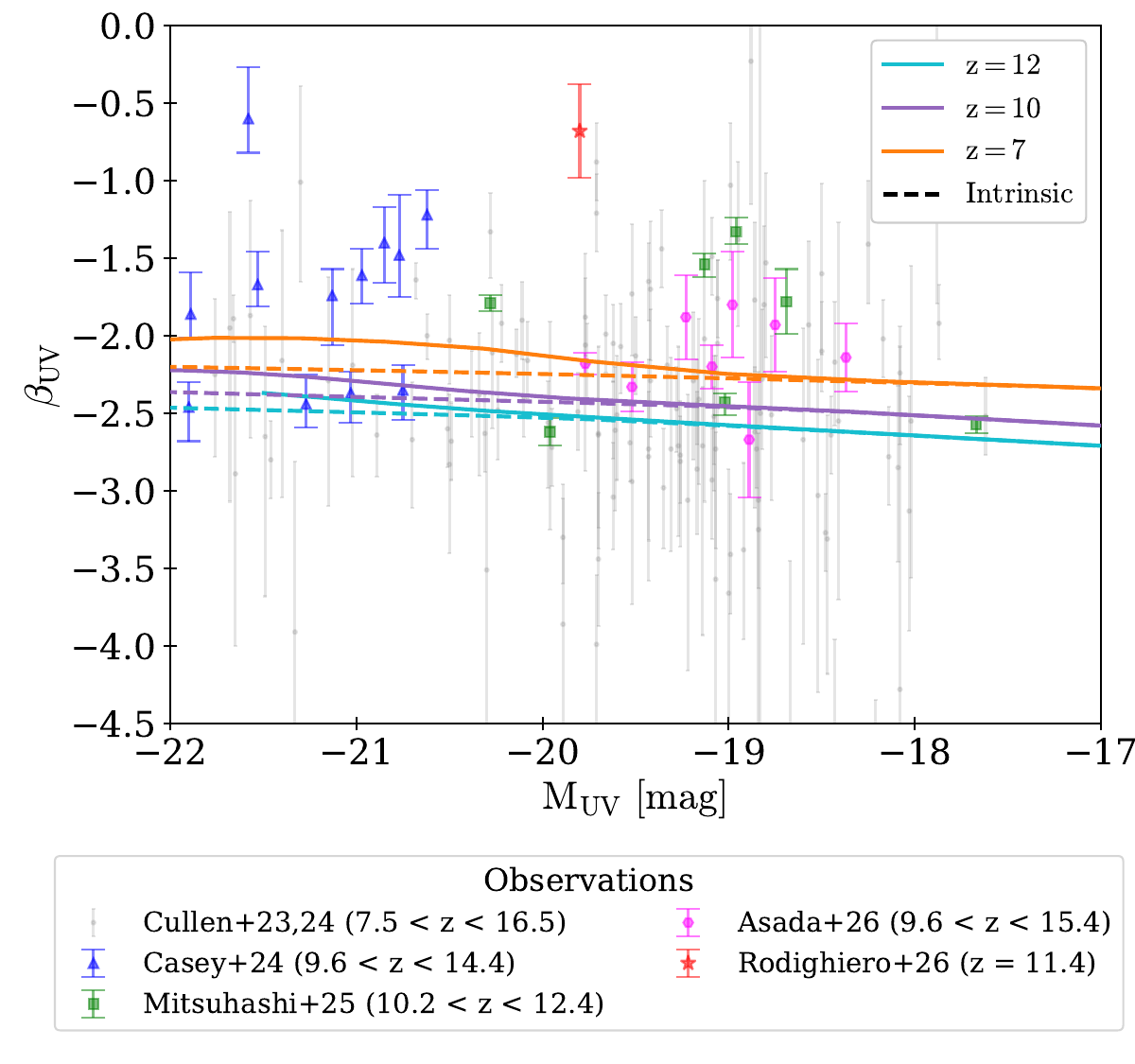}
\caption{
UV slope $\beta_{\rm UV}$ as a function of UV magnitudes $M_{\rm UV}$ of galaxies predicted by our fiducial model is shown by the solid curves,
For comparison, the dashed curves show the intrinsic $\beta_{\rm UV}$--$M_{\rm UV}$ relation without dust attenuation.
The cyan, purple, and orange curves represent the results at $z = 12$, 10, and 7, respectively.
Observational data for individual galaxies at $z > 6$ are shown as 
gray dots \citep{Cullen2023MNRAS, Cullen2024MNRAS}, 
blue triangles \citep{Casey2024ApJ}, 
green squares \citep{Mitsuhashi2025arXiv}, 
magenta circles \citep{Asada2026ApJ}, 
and red stars \citep{Rodighiero2026arXiv}.
}
\label{fig:betauv}
\end{figure}

\begin{figure*}
\centering
\includegraphics[width=1.8\columnwidth]{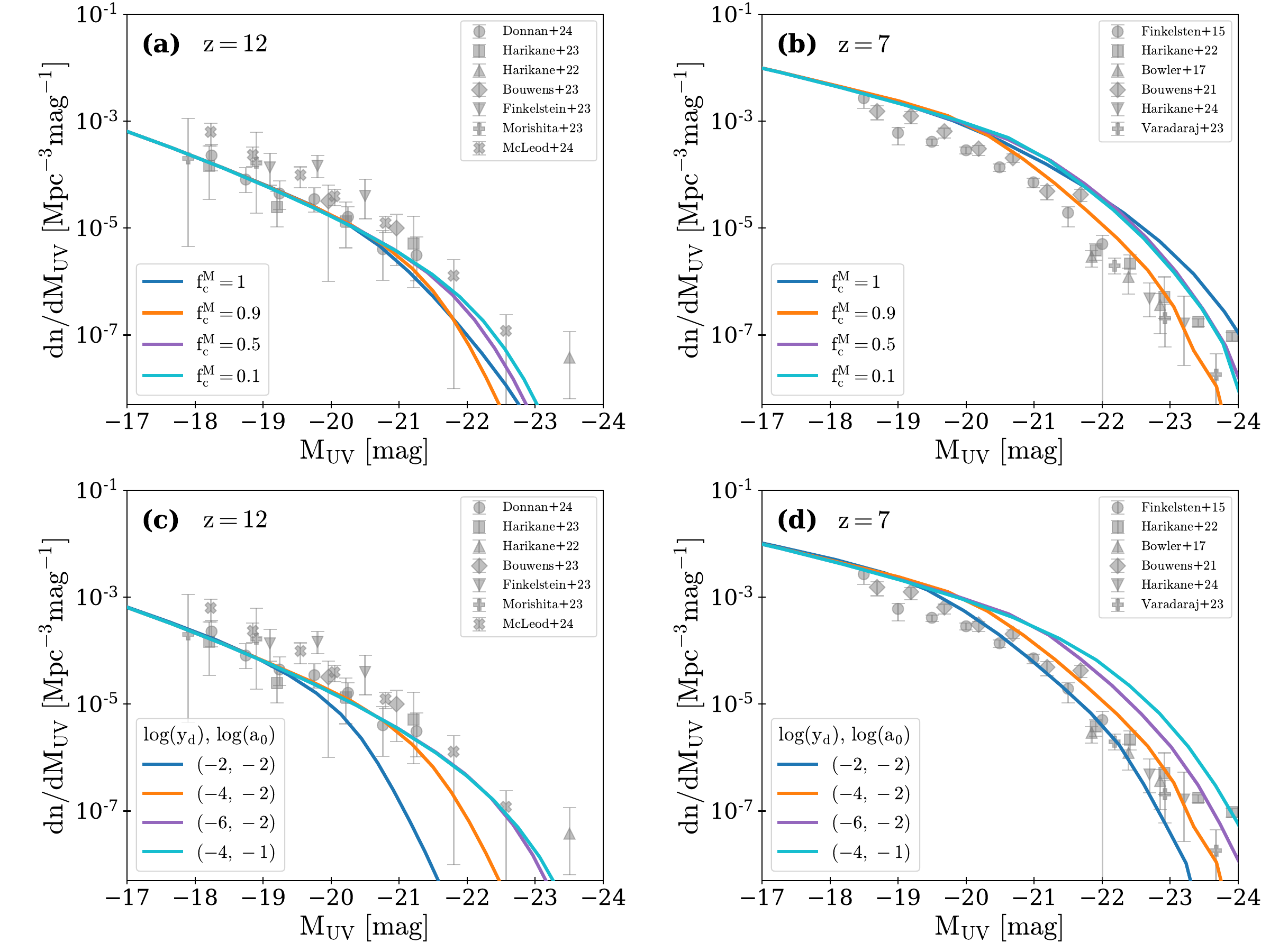}
\caption{
Left and right panels show the UV luminosity functions at $z = 12$ and $z = 7$, respectively, obtained from our model.
Panels (a) and (b) compare models with different CMG mass fractions, $\fcm = 1$ (blue), $0.9$ (orange), $0.5$ (purple), and $0.1$ (cyan).
Panels (c) and (d) compare models with different SN dust properties, $(\log(\yd/\msun),\,\log(a_0/\mu{\rm m})) = (-2,\,-2)$ (blue), $(-4,\,-2)$ (orange), $(-6,\,-2)$ (purple), and $(-4,\,-1)$ (cyan).
In both the upper and lower panels, the orange lines represent the fiducial case in Table~\ref{table:param}.
For comparison, gray symbols show observational data.
Data at $z = 12$ are taken from 
\citet{Harikane2022bApJ, Harikane2023ApJS}, \citet{Bouwens2023MNRAS}, \citet{Finkelstein2023ApJ},
\citet{Morishita2023ApJ}, \citet{Donnan2024MNRAS}, and \citet{McLeod2024MNRAS},
while data at $z = 7$ are taken from 
\citet{Finkelstein2015ApJ}, \citet{Bowler2017MNRAS}, \citet{Bouwens2021AJ}, \citet{Harikane2022aApJS, Harikane2024aApJ}, and \citet{Varadaraj2023MNRAS}.
}
\label{fig:UV_LF}
\end{figure*}

\begin{figure}
\centering
\includegraphics[width=0.9\columnwidth]{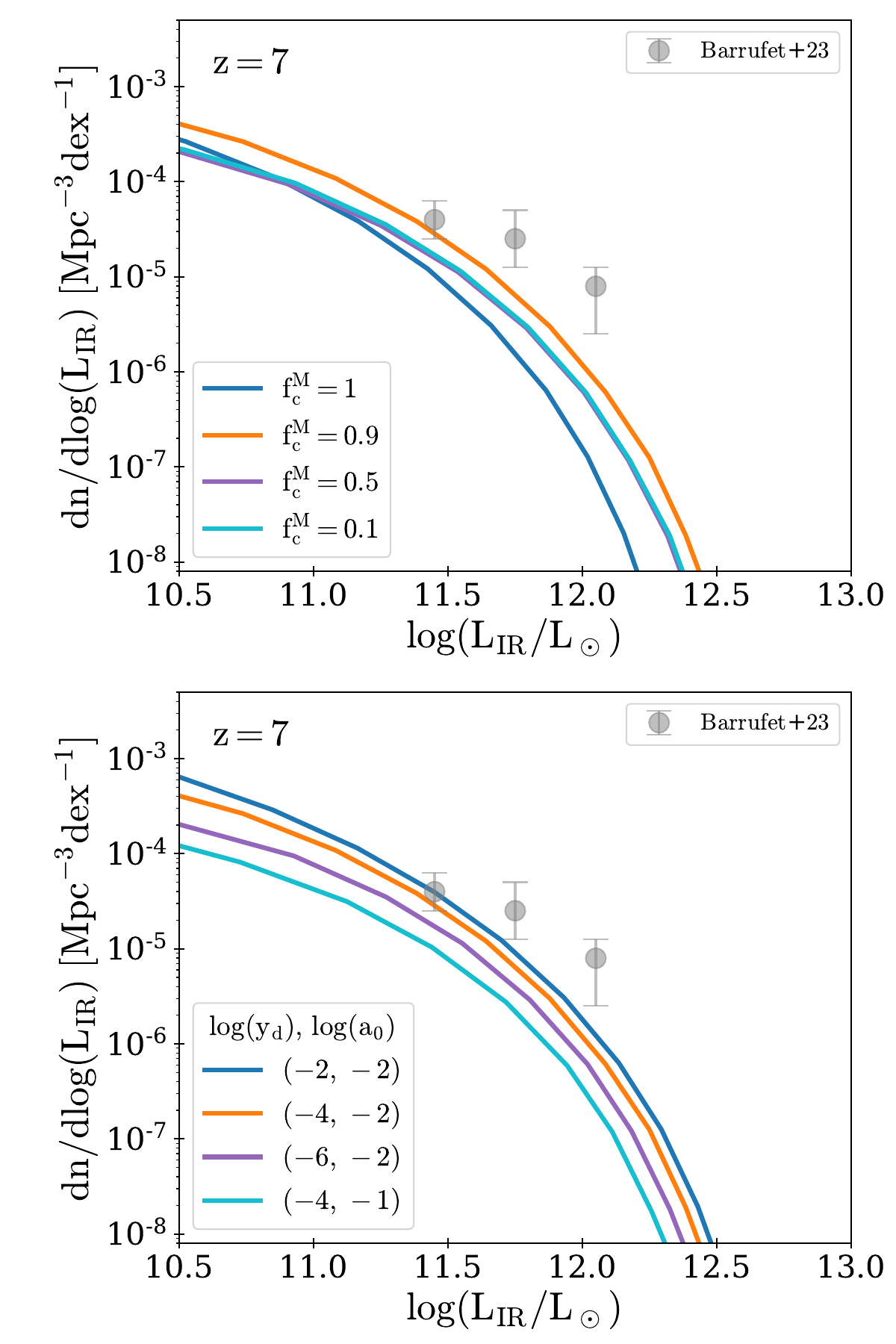}
\caption{
The upper panel shows the IR luminosity functions at $z = 7$ for models with different CMG mass fractions, $\fcm = 1$ (blue), $0.9$ (orange), $0.5$ (purple), and $0.1$ (cyan).
The lower panel shows the results for models with different SN dust properties, $(\log(\yd/\msun),\, \log(a_0/\mu{\rm m})) = (-2,\,-2)$ (blue), $(-4,\,-2)$ (orange), $(-6,\,-2)$ (purple), and $(-4,\,-1)$ (cyan).
Gray symbols show observational data from \citet{Barrufet2023MNRAS}.
}
\label{fig:IR_LF}
\end{figure}

\subsection{UV escape fraction and UV-slope}\label{sec:fesc}

Figure~\ref{fig:fesc} shows the escape fraction of UV photons ($f^{\rm UV}_{\rm esc}$) from galaxies as a function of stellar mass at different redshifts, obtained from our fiducial model.
We find a negative correlation between the two quantities that is nearly independent of redshift.
Low-mass galaxies with $M_\ast \lesssim 10^{8}~\msun$ remain highly transparent, exhibiting $f^{\rm UV}_{\rm esc} > 0.9$.
Above this stellar mass, the escape fraction declines rapidly, consistent with the rapid increase in $M_{\rm d}/M_\ast$ shown in Figure~\ref{fig:Md_z}.
The escape fraction eventually converges to $f^{\rm UV}_{\rm esc} \sim 0.3$ at $M_\ast \gtrsim 10^{10}~\msun$.
This result indicates that even the most massive galaxies at $z \sim 5$ are attenuated in the UV by only about 1~mag, despite having high dust-to-stellar mass ratios of $M_{\rm d}/M_\ast \sim 10^{-2}$.
Thus, the overall dust attenuation remains relatively modest. 
This is mainly because dust evolution in galaxies lags behind star formation and metal enrichment, leaving the outer regions beyond the half-SFR radius transparent to UV radiation, as shown in \S~\ref{sec:radial}.

The predicted escape fractions are broadly consistent with observations of massive galaxies with $\log(M_\ast/\msun) \gtrsim 10$ at $z \sim 5.5$ identified by the ALPINE survey \citep{Fudamoto2020A&A}.
At lower stellar masses of $\log(M_\ast/\msun) \sim 8$--$10$, the REBELS survey has reported a large diversity of escape fractions, $f^{\rm UV}_{\rm esc} \sim 0.1$--0.5 at $z \sim 7$ \citep{Bouwens2022ApJ, Inami2022MNRAS}.
Our prediction can reproduce the upper end of the observed distribution.
Galaxies with $f^{\rm UV}_{\rm esc} \lesssim 0.4$ in this stellar mass range may instead reflect observational selection effects.
The REBELS survey mainly targeted UV-bright galaxies, many of which are likely undergoing intense star formation activity possibly triggered by galaxy mergers \citep{Komarova2026ApJ}.
In such systems, tidal compression may produce denser ISM environments than those assumed in our model, thereby promoting more efficient dust growth and leading to lower UV escape fractions.

Next, we discuss the UV slope $\beta_{\rm UV}$.
The solid curves in Figure~\ref{fig:betauv} present $\beta_{\rm UV}$ as a function of UV magnitude ($M_{\rm UV}$) at different redshifts, predicted by our fiducial model,
whereas the dashed curves show the intrinsic slope $\beta^{\rm int}_{\rm UV}$ without dust attenuation.
The dust-attenuated UV slopes span a range of $-2.5 \lesssim \beta_{\rm UV} \lesssim -2$,
indicating that our model galaxies remain relatively blue across a wide range of $M_{\rm UV}$ and redshifts.

We find that $\beta_{\rm UV}$ gradually increases toward lower redshifts.
This redshift dependence mainly reflects the increasing contribution of older stellar population with cosmic time,
since the same trend persists even in the intrinsic UV slope.
In particular, faint galaxies with $M_{\rm UV} \gtrsim -19$ mag exhibit $\beta_{\rm UV} \sim \beta^{\rm int}_{\rm UV}$ as they remain dust-poor and sufficiently transparent to the UV radiation.
In contrast, galaxies brighter than $M_{\rm UV} \sim -20$ mag at $z \sim 7$ show attenuated UV slopes that are moderately flatter than the intrinsic ones, as efficient dust growth leads to $A_{\rm V} > 0.5$.
However, the deviation remains relatively modest.
This is consistent with the fact that such dust-obscured galaxies have flatter attenuation curves, as shown in Figure~\ref{fig:att_curve}.

In Figure~\ref{fig:betauv}, we compare our result with observations of galaxies at $z > 6$ \citep{Cullen2023MNRAS, Cullen2024MNRAS, Casey2024ApJ, Austin2025ApJ, Mitsuhashi2025arXiv, Asada2026ApJ, Rodighiero2026arXiv}.
The predicted relation between $\beta_{\rm UV}$ and $M_{\rm UV}$ roughly traces the median trend of the observed distribution.
It is worth noting here that observations show a large diversity in UV slope by $\Delta\beta_{\rm UV} \sim 3$.
Furthermore, recent observations report exceptionally red galaxies with $\beta_{\rm UV} > -1.5$ even at $z > 10$ \citep{Mitsuhashi2025arXiv, Rodighiero2026arXiv}.
Such diversity may arise from variations in halo mass assembly histories, which stochastically enhance or suppress dust growth \citep{Di_Cesare2023MNRAS, Narayanan2026OJAp}.
Although reproducing the observed $\beta_{\rm UV}$--$M_{\rm UV}$ relation is beyond the scope of this paper, future work will investigate this issue by incorporating such stochastic effects into our galaxy evolution model.

\subsection{UV and IR luminosity functions}\label{sec:UV_LF}

Here, we comprehensively study UV and IR luminosity functions,
as they are key observables for understanding intrinsic star formation activity and dust attenuation in galaxies.

We first discuss the UV luminosity functions predicted by our model.
We compute the dust-attenuated UV luminosities for 21 dark matter halos spanning $\mhfive = 10^{10}$--$10^{14}~\msun$,
and construct the UV luminosity functions at any redshifts by combining these results with the Sheth–Tormen dark matter halo mass function \citep{Sheth2002MNRAS}.

In Figure~\ref{fig:UV_LF}, we compare the predicted UV luminosity functions with observations of high-redshift galaxies (see the caption for reference).
Panels (a) and (b) show the results at $z = 12$ and $z = 7$, respectively, for four models with different values of $\fcm$.
At $z = 12$, all models are broadly consistent with the observational data within the error bars, indicating that the UV luminosity of galaxies at this early stage is not strongly sensitive to the CMG mass fraction.
At $z = 7$, the fiducial model with $\fcm = 0.9$ successfully reproduces the observed UV luminosity function, whereas the models with $\fcm = 0.1$, $0.5$, and $1$ overpredict the galaxy number density at the bright end. 
This suggests that dust attenuation is particularly efficient in the fiducial model.
Indeed, dust growth is most efficient when the ISM is dominated by CMG while still containing a small fraction of WNG, which provides small grains through shattering, as shown in upper panels of Figure~\ref{fig:Md_z}.
Thus, efficient dust growth in a CMG-rich ISM is favored to reproduce the observed UV luminosity function at $z = 7$.

Panels (c) and (d) of Figure~\ref{fig:UV_LF} compare the UV luminosity functions for four models with different SN dust properties ($\yd,\,a_0$).
In this case, significant differences among the models are already apparent at $z = 12$.
The model with $\yd = 10^{-2}~\msun$ significantly underpredicts the galaxy number density, consistent with the fact that this model tends to overestimate the dust-to-stellar mass ratio at $z > 8$, as shown in lower panels of Figure~\ref{fig:Md_z}.
On the other hand, at $z = 7$, the models with $\yd = 10^{-6}~\msun$ and $a_0 = 0.1~\mu{\rm m}$ overpredict the galaxy number density, reflecting the delayed dust growth in these models.

Overall, the fiducial parameter set, $\yd \sim 10^{-4}~\msun$ and $a_0 \sim 0.01~\mu{\rm m}$, provides a good match to the UV luminosity functions observed at both $z = 7$ and $z = 12$.
These inferred SN dust properties are consistent with recent theoretical predictions that include dust destruction by reverse shocks in the ISM \citep[][]{Otaki2026A&A}.
Thus, the UV luminosity functions of high-redshift galaxies ($z > 5$) provide important constraints on the structure of the multiphase ISM and the properties of SN-produced dust through dust attenuation.

Next, we investigate IR luminosity functions.
Figure~\ref{fig:IR_LF} compares the predicted luminosity function at $z = 7$ with the REBELS observations \citep{Barrufet2023MNRAS}.
The upper and lower panels show the dependence on $\fcm$ and on the SN dust properties ($\yd,\,a_0$), respectively.
As a natural consequence of energy conversion from UV to IR radiation, 
models that predict lower number densities of UV-bright galaxies tend to produce larger populations of IR-bright galaxies.
We find that most of our models significantly underpredict the observed galaxy number density inferred from the REBELS survey.
Although the fiducial model and the model assuming a high SN yield of $\yd = 0.01~\msun$ reproduce the observation around ${\rm log}(L_{\rm IR}/L_\odot) \sim 11.5$, they still underpredict the abundance of galaxies at the brighter side.

This result consistently reflects the tendency of our model tend to predict higher UV escape fractions than those inferred for REBELS galaxies (see Figure~\ref{fig:fesc}).
As already discussed in \S~\ref{sec:fesc}, the REBELS survey mainly targeted UV-bright galaxies, many of which are likely undergoing dusty starburst episodes.
This implies that the REBELS galaxy sample may have systematically different UV and IR radiative properties from the galaxy samples used to derive the UV luminosity functions at $z = 7$.
Such selection effects should therefore be considered more carefully when comprehensively comparing our model predictions with the observed UV and IR luminosity functions in future work.


\section{Discussion}\label{sec:discuss}

\subsection{Comparison with other theoretical studies}

Our model successfully reproduces a broad range of observed properties of galaxies at $z > 5$, including the dust-to-stellar mass ratios and the UV luminosity functions.
Here, we discuss the implications of these results for recent theoretical studies of dust attenuation in high-redshift galaxies.

We first compare the current model with our previous semi-analytic model presented in \citetalias{Toyouchi2025MNRAS}.
In that study, dust growth in the ISM was modeled using a simplified metal accretion timescale that depends only on metallicity as $\tau_{\rm acc} = 5~\Myr~(Z/Z_\odot)^{-1}$.
Even with this simplified treatment, the model successfully reproduced the observed dust-to-stellar mass ratios of galaxies at $z > 5$, including recent constraints from \citet{Bakx2026MNRAS}.
However, the predicted dust attenuation was too strong: galaxies at the bright end of the UV luminosity function experienced $\sim 2$ mag and $\sim 4$ mag of UV attenuation at $z = 12$ and $7$, respectively, causing the model to severely underpredict the abundance of UV-bright galaxies.

In contrast, the current model, which explicitly follows grain-size evolution, successfully reproduces the observed UV luminosity function at both $z = 12$ and $7$.
The primary difference from \citetalias{Toyouchi2025MNRAS} is that the metal accretion timescale now depends explicitly on gas density.
As a result, dust growth proceeds preferentially in the dense central regions of galaxies, while the outer disk regions remain relatively optically thin to UV radiation.
Consequently, galaxies can maintain high UV escape fractions even after becoming substantially dust rich.
Our results therefore demonstrate that the radial distribution of dust, rather than the total dust mass alone, plays a crucial role in determining the UV luminosities of galaxies at cosmic dawn.

This conclusion contrasts with the assumptions adopted in the semi-analytic model of \citet{Mauerhofer2023MNRAS}, who assumed dust distributions extending several times more broadly than the stellar distributions.
Such extended dust geometries significantly reduce the effective optical depth, thereby allowing galaxies to remain UV bright even after becoming dust rich.
A similar possibility has also been discussed by \citet{Ferrara2023MNRAS} and related studies \citep{Fiore2023ApJ, Ziparo2023MNRAS, Nakazato2025MNRAS}, in which stellar feedback temporarily ejects dust from star-forming regions and transports it to the outskirts of galaxies.
Since our current model does not explicitly include dust ejection or dust transport processes, we cannot rule out the importance of such mechanisms.
We discuss the possible impact of dust redistribution in more detail in the next subsection.

Our model further demonstrates that small dust yields of $\yd \leq 10^{-4}~\msun$ is favored to reproduce the extended bright end of the UV luminosity function at $z = 12$, as shown in Figure~\ref{fig:UV_LF}.
Interestingly, models with higher dust yields fail to explain the observational upper limits on dust masses ratios of several galaxies at $z > 8$.
Such low dust yields may naturally arise from efficient dust destruction by reverse shock in dense ISM environments, as predicted by \citet{Otaki2026A&A}.

Recently, \citet{Sommovigo2026arXiv} modeled the UV luminosity functions of high-$z$ galaxies assuming constant dust yields and 
showed that observations at $z \sim 12$ require dust yields of $\yd \sim 0.01~\msun$, which correspondingly gives a dust-to-stellar mass ratio of $M_{\rm d}/M_\ast \sim 10^{-4}$.
This dust yield is about two orders of magnitudes higher than our inferred value.
This discrepancy arises primarily because their model does not include dust growth in the ISM.
Indeed, our model predicts that, even at $z = 12$, dust growth efficiently increases dust-to-stellar mass ratios to $M_{\rm d}/M_\ast \sim 10^{-3}$ in galaxies with $M_\ast \sim 10^{9}~\msun$.

Such rapid dust growth results from the assumed small characteristic size of SN-produced dust, $a_0 = 0.01~\mu$m, 
which efficiently provides seeds for metal accretion.
Importantly, such small stellar grain sizes are themselves a natural consequence of strong reverse shock, together with the reduced dust yield \citep{Otaki2026A&A}.
This implies that intense dust destruction is not necessarily disadvantageous for dust growth in high-$z$ galaxies.
Namely, although reverse shocks reduce the total amount of surviving stellar dust, they can simultaneously promote subsequent grain growth in the ISM by supplying smaller grains with larger surface-area-to-volume ratio.

\citet{Sommovigo2026arXiv} also demonstrated that even with $\yd \sim 0.01~\msun$, dust attenuation remains too weak, causing their model to overpredict the number density of UV-bright galaxies at $z = 7$.
In contrast, our model successfully reproduces the observations at $z = 7$ when adopting a high CMG mass fraction of $\fcm = 0.9$, which efficiently promotes dust growth.
These results indicate the importance of properly modeling the multi-phase structure of the ISM and its role in regulating dust growth..

Our results are also broadly consistent with recent cosmological galaxy formation simulations incorporating grain-size evolution by \citet{Narayanan2026OJAp}.
They found that dust growth proceeds preferentially in dense regions of galaxies and that the dust mass rapidly increases once galaxies reach stellar masses of $M_\ast \sim 10^8$--$10^9~\msun$, in good agreement with our results shown in Figure~\ref{fig:Md_z}.
Furthermore, they argued that galaxies at $z > 10$ exhibit extremely top-heavy grain-size distributions, producing flat attenuation curves and consequently very blue UV spectra.
This trend is also broadly consistent with our predictions shown in Figures~\ref{fig:hist1} and \ref{fig:att_curve}.

There are, however, subtle differences between the two studies.
\citet{Narayanan2026OJAp} argued that dust growth in the ISM remains inefficient at $z > 10$, implying that the top-heavy grain-size distributions largely preserve the initial properties of stellar dust produced by SNe II.
In contrast, our model predicts that dust growth is already efficient even at $z > 10$, such that grain-size distributions naturally evolve toward large-grain-dominated states regardless of the assumed initial stellar dust properties.
One possible reason for this discrepancy is that their simulations adopted intrinsically more top-heavy stellar dust size distributions than those assumed in our fiducial model, making direct comparisons non-trivial.

Overall, these comparisons suggest that rapid dust growth in dense ISM regions, together with centrally concentrated dust distributions and the emergence of large-grain-dominated opacity, are likely key ingredients for understanding the radiative properties of galaxies during cosmic dawn.

\subsection{Implications for attenuation-free scenario}\label{sec:caveats}

Based on our results, we here revisit the attenuation-free scenario, which was originally proposed by \citet{Ferrara2023MNRAS} to explain the extremely blue colors of galaxies at $z > 10$ \citep[e.g.,][]{Fiore2023ApJ, Ziparo2023MNRAS, Nakazato2025MNRAS}.
This scenario supposes that radiation pressure (RP) associated with intense star formation temporarily ejects dust from star-forming regions.
This process effectively reduces the optical depth toward UV-emitting stars, thereby explaining both the bright UV luminosities and non-detection of dust thermal emission commonly reported in galaxies at $z > 10$.

Our current model already includes dust outflows driven by SN-feedback, which can be physically distinct from RP-driven winds.
SN-driven outflows are expected to operate in a relatively continuous manner, whereas RP-driven winds intermittently occur only when specific star formation rates exceed a critical value \citep{Ferrara24}.
As a result, the mass-loading factors, terminal velocities, and dust transport efficiencies may differ substantially between the two mechanisms.
In particular, a significant fraction of dust accelerated by RP-driven winds is expected to remain gravitationally bound to the host galaxy, thereby re-accreting at later times.

If such recycling processes are incorporated, the effective dust mass responsible for UV attenuation at $z > 10$ would temporarily decrease. 
Although the amount of re-accreted dust may be limited,
the recycled dust could still serve as seeds for subsequent grain growth in the ISM.
Consequently, the total dust content of galaxies at later epochs such as $z \sim 7$ may remain largely unchanged.
As shown in Figure~\ref{fig:Md_z}, our current predictions for the dust-to-stellar mass ratios at $7 < z < 12$ are broadly consistent with existing observational constraints \citep[e.g.,][]{Algera2023MNRAS, Bakx2025MNRAS, Bakx2026MNRAS}.
However, future observations may place even tighter upper limits on dust masses at $z > 10$.
If so, temporary dust ejection could become an important mechanism for reconciling theoretical models with observations.

Dust ejection may also affect the local dust geometry within galaxies.
In our model, the strong central concentration of dust produces heavily obscured inner regions, resulting in prominent UV-dark holes inside the half-SFR radius (Figure~\ref{fig:RP}).
However, a certain fraction of observed dusty galaxies do not exhibit such clearly defined UV-dark central regions.
If small-scale dust ejection locally creates spatial segregation between stars and dust even within the central regions of galaxies, a non-negligible fraction of UV photons may escape from otherwise optically thick environments \citep[e.g.,][]{Thompson2016MNRAS, Kim2018ApJ, Menon2025ApJ}.
Such local dust redistribution may therefore explain the observed diversity in the centra UV morphologies of dusty galaxies.

In addition, dust ejection may influence the global spatial distribution of dust across galactic disks.
Several simulations have shown that gas and dust expelled from galactic centers by stellar feedback can gain angular momentum and preferentially re-accrete onto the outer regions of disks \citep[e.g.,][]{Bekki2009ApJ, Gibson2013A&A}.
As shown in Figure~\ref{fig:Reff}, our model consistently predicts compact IR emission with effective radii smaller than those of UV emission by factors of a few.
While this trend agrees with observations of massive galaxies, it fails to explain relatively low-mass galaxies with $\log(M_\ast/\msun) \lesssim 10.5$, which often exhibit extended dust thermal emission with $R_{\rm e,IR}/R_{\rm e,UV} > 1$ \citep{Pozzi2024A&A, Mitsuhashi2024A&A}.
If large-scale galactic fountains preferentially operate in such low-mass systems owing to their shallower gravitational potentials, 
the observed extended IR emission may naturally arise.

Overall, these considerations suggest that dust redistribution processes, including dust ejection, and re-accretion, may play an important role in shaping the radiative properties of high-redshift galaxies.
In future work, we plan to incorporate such processes self-consistently together with grain-size evolution and multiphase ISM physics, enabling a more comprehensive investigation of the spatial distributions of stars, metals, and dust during cosmic dawn.

\section{Summary}\label{sec:summary}

In this paper, we developed a dust evolution model for high-redshift galaxies by combining the galaxy evolution model of \citet{Toyouchi2025MNRAS} with a grain-size evolution model in a two-phase ISM consisting of a cold molecular gas (CMG) component and a warm neutral gas (WNG) component.
With this framework, we followed the coupled evolution of dust mass, grain size distribution, extinction/attenuation curves, and the UV/IR luminosities of galaxies from $z=20$ to $z=5$.
Our results demonstrate that dust growth in high-$z$ galaxies is regulated by both the multiphase ISM structure and the properties of SN-produced dust.
Our main findings are summarized as follows:

\begin{enumerate}
    \item Dust growth becomes most efficient when the ISM is dominated by CMG but still contains a small amount of WNG ($\fcm \sim 0.9$), because CMG promotes metal accretion while WNG supplies small grains through shattering that act as seeds for subsequent grain growth.
    
    \item We adopted a low dust yield of $\yd = 10^{-4}~\msun$ and a small characteristic size of stellar dust of $a_0 = 0.01~\mu$m, motivated by efficient dust destruction by reverse shocks in dense ISM environments. Even with such low dust yields, massive galaxies with $M_\ast > 10^9~\msun$ reach high dust-to-stellar mass ratios of $M_{\rm d}/M_\ast \sim 10^{-2}$ by $z \sim 7$. This rapid dust growth occurs because small grains supplied by SNe efficiently serve as seeds for metal accretion in the ISM. The fiducial model adopting $\fcm \sim 0.9$, $\yd = 10^{-4}~\msun$, and $a_0 = 0.01~\mu$m successfully reproduces the observed dust and stellar masses of galaxies at $z = 6$-14.
    
    \item The grain size distribution evolves rapidly in the dense inner regions of galaxies.
    Even if stellar dust is initially dominated by small grains with $a \sim 0.01~\mu{\rm m}$, metal accretion in the ISM leads to a large fraction of grains with $a \gtrsim 0.1~\mu{\rm m}$ by $z \sim 7$ in massive galaxies.
    As a result, extinction curves become relatively flat in the inner regions, while the outer regions retain steeper, small-grain-dominated extinction curves.

    \item The predicted attenuation curves become flatter with increasing $A_{\rm V}$. In addition, the attenuation curves tend to be flatter than the corresponding mass-weighted extinction curves because dust attenuation predominantly occurs in the inner regions, where the grain size distribution is strongly biased toward large grains.

    \item Dust growth significantly lags behind metal enrichment in the outer parts of galaxies.
    Consequently, dust is much more centrally concentrated than metals and stars, and a non-negligible fraction of the escaping UV radiation originates from the relatively dust-poor outer disk.
    This naturally reduces the effective dust attenuation of galaxies, even when their central regions are already highly dust-rich.

    \item Our model predicts that dust thermal emission is systematically more compact than UV emission.
    This prediction is broadly consistent with observations of massive dusty galaxies, although the extended IR emission observed in less massive systems may require additional processes such as dust ejection and subsequent re-accretion, which constitute key ingredients of attenuation-free scenarios.
    
    \item The UV luminosity functions at $z=7$ and $z=12$ provide strong constraints on dust physics in high-$z$ galaxies.
    The fiducial model successfully reproduces the observed UV luminosity functions, whereas models with too small CMG fraction ($\fcm < 0.5$), 
    excessively large SN dust yields ($\yd \sim 10^{-2}~\msun$), or large initial grain sizes ($a_0 \sim 0.1~\mu{\rm m}$) fail to reproduce both epochs simultaneously.
    On the other hand, we note that our models underestimate the IR luminosity function at $z \sim 7$ relative to that inferred by the REBELS survey.

\end{enumerate}

Overall, this study demonstrates that a physically motivated treatment of grain growth in a multiphase ISM is essential for linking the dust content of high-redshift galaxies to their radiative properties during cosmic dawn.

Our theoretical framework can also be extended to enable more comprehensive comparisons with future multi-wavelength observations of high-redshift galaxies.
In future work, we plan to calculate the spatial distribution of dust temperatures by considering the non-uniform density structure of the multiphase ISM.
This will allow us to construct self-consistent spectral energy distributions spanning from the UV to the far-IR.
Using such calculations, we will investigate both the radiative properties of galaxies observed at $z < 8$ and the detectability of dust thermal emission from galaxies at $z > 10$ in a unified manner.

The present study also suggests several important directions for improving theoretical models themselves.
In particular, it will be important to incorporate dust ejection and subsequent re-accretion in order to investigate the spatial redistribution of dust in galaxies.
In addition, as some high-redshift galaxy samples may be observationally biased toward UV-bright phases, potentially associated with galaxy mergers. 
Accounting for such intermittent starburst events in our model will therefore be crucial for fair comparisons with observations.
By combining these effects with grain-size evolution and multiphase ISM physics, future models will provide a more complete understanding of the coupled formation of stars, metals, and dust during the earliest phases of galaxy formation.



\section*{Acknowledgements}

We thank K.~Nagamine, R.~Tazaki, and A.~Lupi for fruitful discussions.
We are grateful to S.~Fujimoto for kindly providing the observational data from \citet{Fujimoto2017ApJ}.
D.~T.\ was supported in part by JSPS KAKENHI Grant Number JP25K17438.
This work was also supported by the JSPS International Leading Research (ILR) program, JP22K21349.
AF acknowledges support from the ERC Advanced Grant INTERSTELLAR H2020/740120. 
This research was supported (AF) in part by grant NSF PHY-2309135 to the Kavli Institute for Theoretical Physics (KITP). 
YN acknowledges Flatiron Research Fellowhship. The Flatiron Institute is a division of the Simons Foundation. 


\section*{Data availability}

The data underlying this article will be shared on reasonable request to the corresponding author.




\bibliographystyle{mnras}
\bibliography{refs.bib} 




\bsp	
\label{lastpage}
\end{document}